\documentclass[3p]{elsarticle}

\usepackage{lineno}
\usepackage{hyperref}
\usepackage{caption}
\usepackage{subcaption}
\modulolinenumbers[5]

\usepackage{amsmath}
\usepackage{multirow}
\usepackage{graphicx}
\usepackage{float}
\usepackage{xcolor}
\usepackage[linesnumbered,ruled,vlined]{algorithm2e}
\usepackage{adjustbox}
\RestyleAlgo{ruled}
\SetKwComment{Comment}{/* }{ */}
\usepackage{tabularx}
\usepackage{rotating}

\usepackage{hyperref}

\usepackage[nolist]{acronym}
\begin{acronym}[AoI]
\acro{AoI}{Area of Interest}
\end{acronym}
\begin{acronym}[ARQ]
\acro{ARQ}{Automatic Repeat Request}
\end{acronym}
\begin{acronym}[CAM]
\acro{CAM}{Cooperative Awareness Message}
\end{acronym}
\begin{acronym}[CBR]
\acro{CBR}{Channel Busy Rate}
\end{acronym}
\begin{acronym}[ETSI]
\acro{ETSI}{European Telecommunications Standards Institute}
\end{acronym}
\begin{acronym}[ITS]
\acro{ITS}{Intelligent Transport Systems}
\end{acronym}
\begin{acronym}[DCC]
\acro{DCC}{Decentralized Congestion Control}
\end{acronym}
\begin{acronym}[CA]
\acro{CA}{Cooperative Awareness}
\end{acronym}
\begin{acronym}[CAM]
\acro{CAM}{Cooperative Awareness Message}
\end{acronym}
\begin{acronym}[CBF]
\acro{CBF}{Contention-Based Forwarding}
\end{acronym}
\begin{acronym}[DEN]
\acro{DEN}{Decentralized Environmental Notification}
\end{acronym}
\begin{acronym}[DENM]
\acro{DENM}{Decentralized Environmental Notification Message}
\end{acronym}
\begin{acronym}[DPD]
\acro{DPD}{Duplicate Packet Detection}
\end{acronym}
\begin{acronym}[DPL]
\acro{DPL}{Duplicate Packet List}
\end{acronym}
\begin{acronym}[GPC]
\acro{GPC}{Geographically-aware CBF Packet Cancellation}
\end{acronym}
\begin{acronym}[FCD]
\acro{FCD}{Floating Car Data}
\end{acronym}
\begin{acronym}[FoT]
\acro{FoT}{Forward on Time}
\end{acronym}
\begin{acronym}[FoT+]
\acro{FoT+}{Forward on Time+}
\end{acronym}
\begin{acronym}[GBC]
\acro{GBC}{Geographically-Scoped Broadcast}
\end{acronym}
\begin{acronym}[IVI]
\acro{IVI}{Infrastructure to Vehicle Information}
\end{acronym}
\begin{acronym}[IVI]
\acro{IVIM}{IVI Message}
\end{acronym}
\begin{acronym}[C-ITS]
\acro{C-ITS}{Cooperative Intelligent Transport Systems}
\end{acronym}
\begin{acronym}[LocT]
\acro{LocT}{Location Table}
\end{acronym}
\begin{acronym}[LocTE]
\acro{LocTE}{Location Table entry}
\end{acronym}
\begin{acronym}[PDR]
\acro{PDR}{Packet-delivery Ratio}
\end{acronym}
\begin{acronym}[RSU]
\acro{RSU}{Road-side Unit}
\end{acronym}
\begin{acronym}[RHW]
\acro{RHW}{Road Hazard Warning}
\end{acronym}
\begin{acronym}[S-CBF]
\acro{S-CBF}{Slotted CBF}
\end{acronym}
\begin{acronym}[SHB]
\acro{SHB}{Single-Hop Broadcast}
\end{acronym}
\begin{acronym}[TC]
\acro{TC}{Traffic Class}
\end{acronym}
\begin{acronym}[TRC]
\acro{TRC}{Transmit Rate Control}
\end{acronym}
\begin{acronym}[TTL]
\acro{TTL}{Time-to-Live}
\end{acronym}
\begin{acronym}[VANET]
\acro{VANET}{Vehicular ad hoc Network}
\end{acronym}
\begin{acronym}[VAM]
\acro{VAM}{VRU Awareness Message}
\end{acronym}
\begin{acronym}[VRU]
\acro{VRU}{Vulnerable Road User}
\end{acronym}
\begin{acronym}[VBS]
\acro{VBS}{VRU awareness}
\end{acronym}
\begin{acronym}[V2X]
\acro{V2X}{Vehicle-to-everything}
\end{acronym}
\journal{Computer Communications}

\begin{document}

\begin{frontmatter}

\title{Evaluation of Greedy and CBF for ETSI non-area GeoNetworking: the impact of DCC} 

\tnotetext[mytitlenote3]{\textbf{Published as: Oscar Amador;  Maria Calderon; Manuel Urueña; Ignacio Soto. Evaluation of Greedy and CBF for ETSI non-area GeoNetworking: The impact of DCC. Computer Communications 218, pp.114-130, 2024.  The final version of record is available at \href{https://doi.org/10.1016/j.comcom.2024.02.009}{https://doi.org/10.1016/j.comcom.2024.02.009}}}
\tnotetext[mytitlenote]{This work was partially supported by the Agencia Estatal de Investigaci\'on (AEI, Spain) through the ACHILLES project (PID2019-104207RB-I00/AEI/10.13039/501100011033)}
\tnotetext[titlenote2]{We gratefully acknowledge support from the Swedish Knowledge Foundation (KKS) "Safety of Connected Intelligent Vehicles in Smart Cities -- SafeSmart" project (2019--2024), and the ELLIIT Strategic Research Network.}

\author[Halmstad]{Oscar Amador}
\ead{oscar.molina@hh.se}

\author[UPM]{Maria Calderon}
\ead{maria.calderon@upm.es}

\author[UNIR]{Manuel Urue\~na}
\ead{manuel.uruena@unir.net}

\author[UPM]{Ignacio Soto\corref{mycorrespondingauthor}}
\cortext[mycorrespondingauthor]{Corresponding author}
\ead{ignacio.soto@upm.es}

\address[Halmstad]{School of Information Technology; Halmstad University; Halmstad 30118; Sweden}
\address[UPM]{Departamento de Ingenier\'{\i}a de Sistemas Telem\'aticos; ETSI Telecomunicaci\'on, Universidad Polit\'ecnica de Madrid; 28040 Madrid (Madrid); Spain}
\address[UNIR]{Escuela Superior de Ingenieros y Tecnolog\'{\i}a, Universidad Internacional de la Rioja; 26006 Logro\~no; Spain}

\begin{abstract}
This paper evaluates the performance of the two ETSI non-area forwarding algorithms in the GeoNetworking specification: Greedy Forwarding and Non-Area Contention-Based Forwarding (CBF). Non-area forwarding occurs when a packet is sent to a geographical Destination Area from a node located outside of this area, e.g., when a vehicle wants to alert of hazardous events to other vehicles located in a distant geographical area. The evaluation has been carried out both in urban and highway scenarios and takes into account the complete ETSI Architecture, including the interaction with the \acf{DCC} mechanism. We have also compared ETSI-defined mechanisms with optimizations found in the literature. Our main findings are that Greedy Forwarding, when combined with \ac{DCC}, is extremely ineffective even with optimizations, and Non-Area CBFs (both ETSI CBF and an optimized version called S-FoT+) outperform Greedy Forwarding both in highway and urban scenarios. 
\end{abstract}

\begin{keyword}
ETSI Intelligent Transport Systems (ITS) \sep Non-area GeoNetworking \sep Decentralized Environmental Notification Message (DENM) \sep Cooperative Awareness Message (CAM) \sep Greedy Forwarding \sep Contention-Based Forwarding (CBF) \sep Duplicate Packet Detection (DPD) \sep Decentralized Congestion Control (DCC)
\end{keyword}

\end{frontmatter}

%\linenumbers

\section{Introduction}
\label{sec:intro}

\acf{C-ITS} aim to improve road safety and traffic efficiency, as well as to offer comfort and infotainment applications to passengers by exchanging messages between vehicles, with the infrastructure, and with other traffic actors. A relevant cooperative service is the dissemination of notifications to warn other vehicles about different traffic incidents \cite{etsiRHS}. These notifications are triggered by a detected event (e.g., ice on the road) and they are usually disseminated by a vehicle (e.g., the vehicle that detected the event) or by an element of the infrastructure (e.g., an infrastructure sensor or a video surveillance camera)\cite{Li2021}. The possible events to be notified are very diverse: an approaching emergency vehicle, a stationary vehicle, roadworks, adverse weather conditions, intersection collision warnings, imminent hazards such as a red light violation, other hazard warnings, or traffic jam warnings. These notifications are usually distributed to vehicles in a geographical area that could be affected by the situation, so that they can react accordingly (e.g., by slowing down or re-routing). 

\begin{figure}[tbh!]
    \centering
    \includegraphics[width=0.8\textwidth]{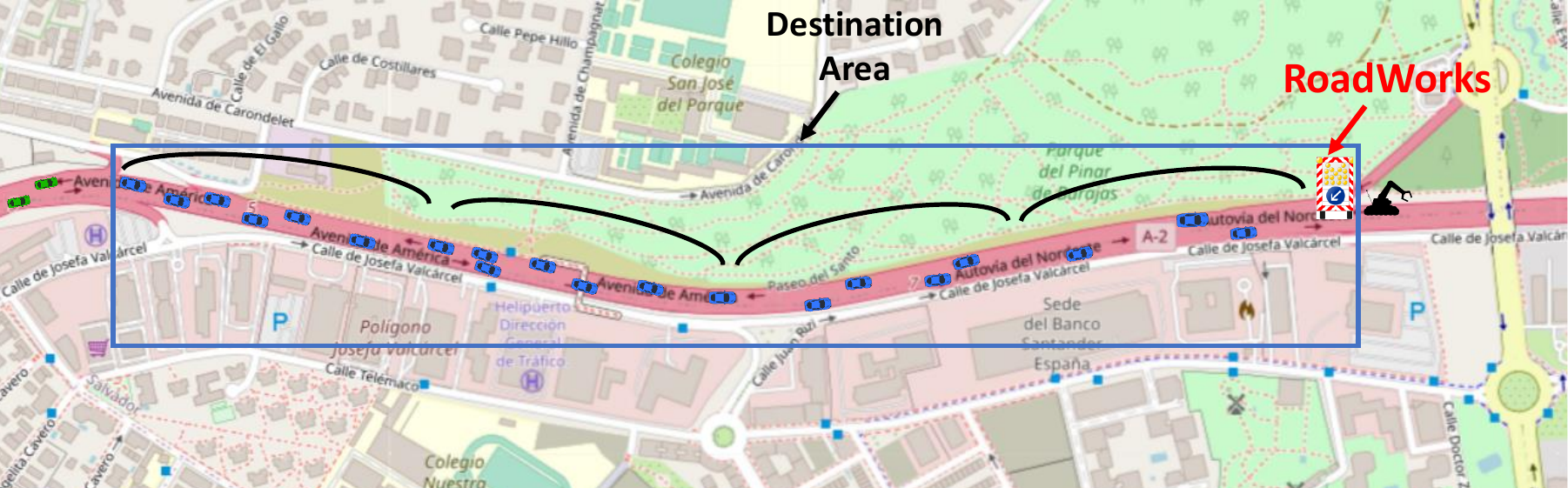}
    \caption{Warning dissemination with the source inside the Destination Area (map from {\href{https://www.openstreetmap.org/copyright}{OpenStreetMap})}}
    \label{fig:area}
\end{figure}

An event-triggered message specifies a Relevance Area (i.e., the area where the event is taking place), and a Destination Area (i.e., the region where the nodes we want to notify are located). Depending on the specific use case, the location of the source (e.g., the node generating the event notification) may vary with respect to the Destination or Relevance Areas (e.g., a source node can notify nodes in a Destination Area of an event happening in a Relevance Area without being in either zone). In the case of road hazard anticipation systems \cite{Ducourthial:2022}, the source may wish to warn approaching vehicles. An example of a road hazard anticipation system is shown in Fig.~\ref{fig:area} where an infrastructure element, such as a road work ahead sign, warns drivers to slow down and be prepared to encounter a change in road conditions. This sign, located in the Destination Area, is the source of notifications to approaching vehicles (i.e., blue cars). The example shows how notification messages may need several hops to cover the complete Destination Area (defined as a rectangle in the figure). 

\begin{figure}[tbh!]
    \centering
    \includegraphics[width=0.8\textwidth]{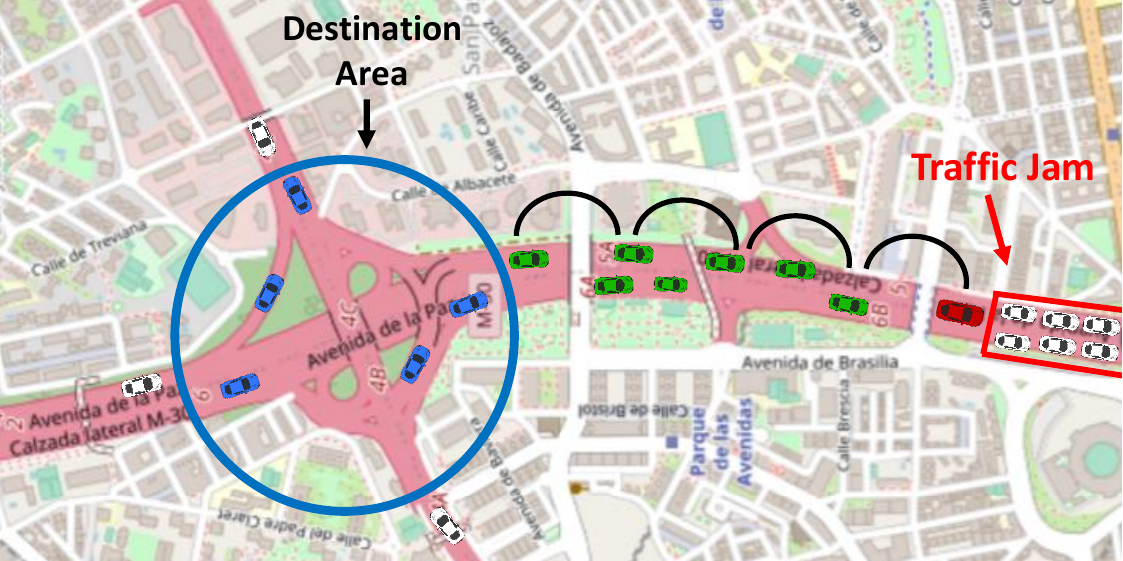}
    \caption{Warning dissemination with the source outside the Destination Area (map from {\href{https://www.openstreetmap.org/copyright}{OpenStreetMap})}}
    \label{fig:non-area}
\end{figure}

Another interesting use case is shown in Fig.~\ref{fig:non-area}, which presents a traffic jam avoidance system with route diversion. This is a relevant use case for traffic efficiency in both urban \cite{xu:2017} and  highways \cite{han2021} environments, considering the time lost every day by drivers due to traffic jams. Fig.~\ref{fig:non-area} shows the case of a vehicle (in red) that detects a traffic jam ahead and warns vehicles in a distant area (circular area in blue) so that they can avoid the traffic jam by changing their driving route (e.g., taking advantage of a nearby junction). In this example, the notification messages will need several hops to reach the Destination Area before being disseminated there. 

The \acf{ETSI} in its specifications for \acf{ITS} includes the Decentralized Environmental Notification (DEN)~\cite{etsiDEN} basic service. This is an event-based message delivery service, which uses \acp{DENM} to alert and convey relevant warning information to affected road users within a Destination Area. \acp{DENM} are the basic building block for \acf{RHW} applications \cite{etsiRHS}. This is one of the first ITS applications to be deployed in real-life scenarios.

The \ac{ETSI} \ac{ITS} specifications \cite{etsiNewGeoNetworking} define mechanisms to disseminate a message inside a Destination Area with a multihop scheme where the message is rebroadcast by vehicles until the complete area is covered. This dissemination inside the Destination Area is regulated by an area forwarding protocol. According to the ETSI specifications, the Destination Area can be rectangular, circular, or ellipsoidal. In case the source wants to disseminate the notification in a distant Destination Area there will be two different steps. In the first step, the message is forwarded toward the Destination Area using a non-area forwarding protocol, and once the message reaches the Destination Area, it is disseminated inside the area, using an area forwarding protocol, and delivered to applications. 

The ETSI GeoNetworking protocol \cite{etsiNewGeoNetworking} is a network layer protocol that controls the forwarding of messages using geographical positions.  Each node maintains a \ac{LocT} with the last known position of the nodes around it. They can be nodes one hop away (neighbors) that have sent a beacon with status information such as their position, or nodes that have distributed a message even if they are several hops away. The \ac{LocT} is used while making forwarding decisions. Several kinds of communications are considered in the GeoNetworking specification: GeoUnicast, to send  messages to individual destinations; \ac{SHB}, for messages addressed to all one-hop neighbors; \textcolor{black} {\ac{GBC}, for messages addressed to all nodes inside a Geographic Destination Area}; multi-hop Topologically-scoped Broadcast (TSB), for messages addressed to n-hop neighbors; and GeoAnycast, for messages addressed to any node inside a Destination Area.

Non-area forwarding uses GeoUnicast communications, and area forwarding uses Geographically-Scoped Broadcast communications. The ETSI GeoNetworking \cite{etsiNewGeoNetworking} protocol specifies two forwarding algorithms for non-area forwarding, these are Greedy Forwarding (the default option) and Non-Area \acf{CBF}. Similarly, other two area forwarding algorithms are defined, namely, Area \ac{CBF} (the default protocol for area forwarding) and Simple area forwarding algorithm (also known as Simple GeoBroadcast). 

The impairments of area forwarding algorithms, Area CBF and Simple area forwarding, have been analyzed thoroughly \cite{Bellache:2017a, Marzouk:2018, Turcanu2020, Spadaccino2020, riebl2021, Hajjej2022, Amador2022, S-FoT+:2023}. These works reveal inefficiencies such as redundant retransmissions due to \acf{DCC} queues \cite{S-FoT+:2023, Kuhlmorgen2020}, the need for persistent duplicate packet detection \cite{riebl2021, Amador2022}, or poor dissemination coverage \cite{Bellache:2017a, Marzouk:2018, Spadaccino2020, S-FoT+:2023, Torrent2007}.  These papers have also proposed a number of enhancements to ETSI Area CBF to overcome the shortcomings identified. However, the non-area forwarding algorithms specified in the ETSI GeoNetworking \cite{etsiNewGeoNetworking} protocol have not received much attention from researchers to date.   

This paper evaluates the performance of ETSI-defined GeoNetworking mechanisms for scenarios where a message must cover a Destination Area and the source node is outside of that area. The contributions of this work are:

\begin{itemize}
    \item An experimental evaluation of ETSI non-area forwarding mechanisms (Greedy Forwarding and Non-Area \ac{CBF}), including a comparison with proposed optimizations selected from the literature.
    \item An evaluation of Greedy Forwarding showing that, even with optimizations to improve its reliability, it performs poorly as non-area forwarding protocol when combined with ETSI \ac{DCC}. 
    \item A validation of optimizations already proposed for Area CBF in non-area scenarios, showing that they outperform ETSI Non-Area CBF in terms of efficiency.
\end{itemize}

The remainder of the paper is organized as follows: Section~\ref{sec:background} describes the ETSI ITS architecture, the forwarding algorithms specified for GeoNetworking, and optimizations proposed in the literature; Section~\ref{sec:evaluation} describes the area and non-area forwarding mechanisms we evaluate in this work; Sections~\ref{sec:highway} and~\ref{sec:urban} present the results of the experimental evaluation of the forwarding mechanisms in highway and urban scenarios, respectively; Section~\ref{sec:relatedWork} presents an exploration of related works assessing non-area forwarding; and finally, Section~\ref{sec:conclusions-fw} presents the conclusions of this work and the future lines of work.

\section{Background} 
\label{sec:background}

In this section, we briefly present the ETSI ITS Architecture. First, the general layout and then the two non-area forwarding algorithms specified in~\cite{etsiNewGeoNetworking} --- Greedy Forwarding and Non-Area \ac{CBF}. Next, we explore two problems identified in the literature affecting the efficacy of Greedy Forwarding and mitigation proposals; and finally, we explore a set of improvements proposed to significantly increase ETSI Area CBF efficiency and whether they apply to Non-Area CBF.

\subsection{ETSI ITS Architecture}

\begin{figure}[tbh!]
    \centering
    \includegraphics[width=0.7\textwidth]{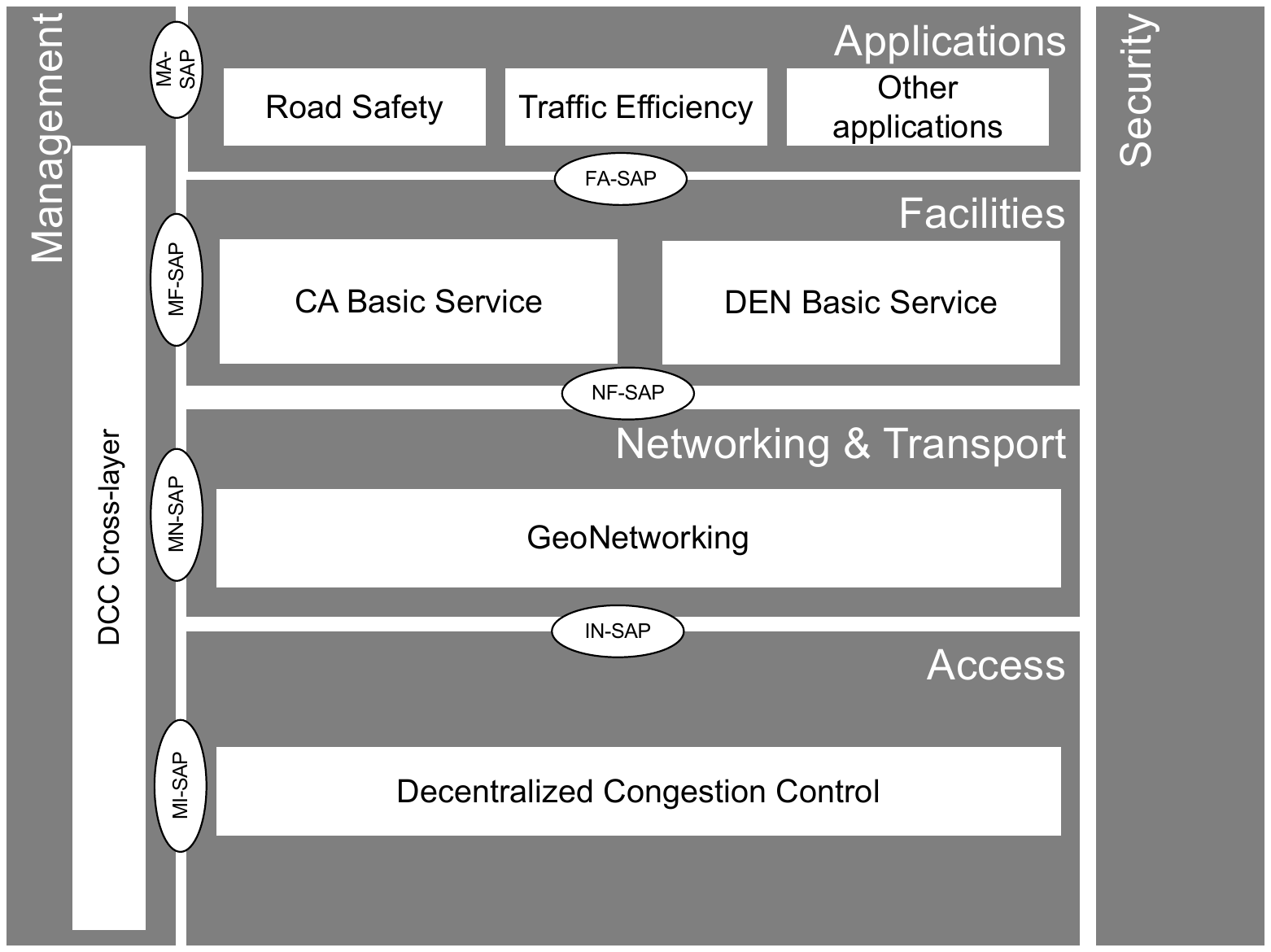}
    \caption{ETSI ITS Architecture}
    \label{fig:architecture}
\end{figure}

\textcolor{black}{The ETSI ITS specification defines a complete architecture in which the ETSI GeoNetworking protocol~\cite{etsiNewGeoNetworking} is placed at the Network \& Transport layer. The Facilities layer leverages the services offered by the GeoNetworking protocol and provides its services to ITS applications. The two most relevant ITS application families are road safety and traffic efficiency, although other applications could also be envisaged (e.g., passenger entertainment). Fig.~\ref{fig:architecture} shows the architecture, highlighting the services that we study in this paper.}

\textcolor{black}{Services that support ITS applications reside in the Facilities layer. These services rely on the exchange of messages with different characteristics and requirements. For example:}
\textcolor{black}{
\begin{itemize}
    \item Services supporting road safety:
    \begin{itemize}
        \item \ac{CA} basic service~\cite{etsiCA}, which uses \acp{CAM}.
        \item Some use cases of the \ac{DEN}~\cite{etsiDEN} basic service --- e.g., emergency electronic braking light (EEBL), stationary vehicle warning --- which are notified using \acp{DENM}.
        \item \ac{VRU} Awareness basic service~\cite{etsi-va}, which uses \acp{VAM} to notify the presence of \acp{VRU} (e.g., pedestrians and cyclists).
    \end{itemize}
    \item Services supporting traffic efficiency:
    \begin{itemize}
        \item \ac{IVI} service~\cite{infrastructureFacilitues}, which uses \acp{IVIM} to disseminate mandatory and advisory road signage.
        \item Some use cases of the \ac{DEN} basic service --- e.g., traffic jam ahead warning.
    \end{itemize}
\end{itemize}
}
\textcolor{black}{From these messages, \acp{CAM} and \acp{VAM} are sent in \ac{SHB}, i.e., they are broadcast to neighbors one hop away from the source. Furthermore, while some \acp{DENM} are only expected to reach immediate neighbors (e.g., in the EEBL use case), some have to reach a destination area that is more than one hop away. This is also the case for some \acp{IVIM}, which require sending messages to a minimum destination area which might exceed the communication range of the source node~\cite{infrastructureFacilitues}. For these cases, \ac{GBC} is employed to reach and cover the destination area.}

In more detail, the Cooperative Awareness service employs \acp{CAM} that are regularly broadcast to one-hop neighbors, delivering status information (e.g., position and heading of the source vehicle). \textcolor{black}{\acp{CAM} are triggered dynamically at intervals between 0.1\,s and 1\,s depending on 1) kinematic thresholds (e.g., changes in position, speed, orientation), and 2) the inter-message rate allowed by the access layer (contingent on the access technology being used).} Furthermore, \acp{DENM} \textcolor{black}{are triggered depending on the use case they solve. They can be triggered through the human-machine interface, or automatically by sensors (e.g., for EEBL). In either case, a \ac{DENM} is generated if a road hazard is recognized, and the message contains information on the nature of the hazard and its location (i.e., the Relevance Area) and, if necessary, the area where vehicles shall learn about the hazard (i.e., the Destination Area).} 

At the Access layer, several technologies are possible in the dedicated 5.9~GHz band, with IEEE 802.11p/ETSI ITS-G5~\cite{etsiMediaDependentG5} being one of the most supported currently. A relevant building block at the Access layer is \acf{DCC}, a mechanism to control the channel load and ensure that the radio medium operates at an efficient regime. 

The architecture allows \ac{DCC} to influence different layers through a DCC cross-layer at the Management entity (see Fig.~\ref{fig:architecture}), e.g., \ac{CAM} generation is controlled by vehicle dynamics, but also restricted by \ac{DCC}. Regarding the DCC mechanism at the Access Layer, an adaptive variant based on the LIMERIC algorithm~\cite{Limeric2013} is specified. Each transmitter independently regulates its message-sending rate (Transmit Rate Control). To this end, the \acf{CBR}, the percentage of time that the channel is busy, is measured and used as an input for a linear control system. This method seeks to collectively converge to a pre-defined level of channel utilization, a target \ac{CBR}. At the Access layer, there are several DCC queues (a DCC queue for each traffic class or priority), and packets could be stored there while waiting to access the channel because of the rate control. The GeoNetworking protocol handles traffic classes to prioritize the network traffic. Each message has an associated \acf{TC} according to its differentiated quality of service. There exist four priorities: TC0 to TC3, with TC3 being the lowest one. CAMs are marked as TC2. As for \acp{DENM}, they have TC0 or TC1 (the highest priorities) at the source node, but they are TC3 when forwarded.

\subsection{Greedy Forwarding}%% Manolo

\begin{figure}[tbh!]
    \centering
    \includegraphics[width=0.8\textwidth]{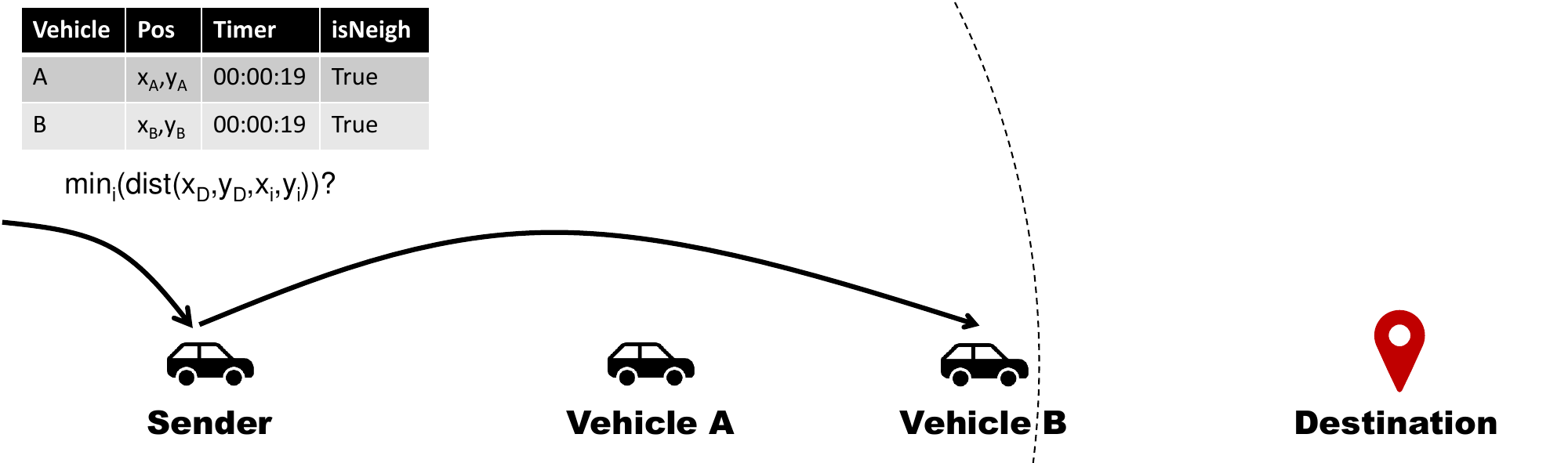}
    \caption{Greedy Forwarding}
    \label{fig:Greedy}
\end{figure}

Greedy Forwarding is the default unicast and non-area forwarding mechanism defined in the ETSI ITS Architecture~\cite{etsiNewGeoNetworking}. 
It is a sender-based forwarding mechanism, where the sender of a packet decides the best next hop toward the destination. This is done by evaluating the progress toward the packet destination, either a single vehicle or the center of a Destination Area. To do so, the sender node first evaluates its distance toward the target coordinates. Then, it looks for all its direct neighbors and evaluates their distances to the target coordinates, ignoring all the neighbors whose distance is greater than its own (to prevent going backward). Then, it chooses the direct neighbor with the shortest distance toward the target destination and selects it as the packet next hop (e.g., Vehicle B in Fig.~\ref{fig:Greedy}).

The packet is then sent in a unicast frame toward the selected next hop. This allows the packet to be re-transmitted several times (up to 8 by default --- one first attempt plus seven retries) in case the previous frames are lost and no acknowledgment frames are received (\ac{ARQ}). However, the specified Greedy Forwarding mechanism does not have any fallback mechanism in case all re-transmissions to this next hop fail, leading to the packet being lost. This reliability issue is acknowledged by the ETSI specification when describing Greedy and Contention-based forwarding in~\cite{etsiNewGeoNetworking}.

When a node receives a packet that is being forwarded using the Greedy Forwarding algorithm, it first checks if it is already on its Duplicated Packet List (DPL) to prevent loops (in which case the packet is just dropped), and if not, it is recorded there and the same process is repeated until the target area or destination coordinates are reached.

If any forwarding node (the sender node) does not find any direct neighbor with better progress than its own, a dead end has been reached. In that case, the ego node just broadcasts the packet once hoping that some unknown neighbor with a better progress receives it and continues its forwarding, or maybe that a worse neighbor does know a better next hop for the packet than the current sender. Given that the packet is broadcast, all vehicles in the vicinity of the sender node will receive it and try to forward it, so it may happen that a single packet is forwarded through different paths after each broadcast transmission, although the duplicate packet list mechanism may also drop all duplicated packets that converge through the same node.

\subsubsection{Stale neighbor information in Greedy Forwarding}

Although the Greedy Forwarding algorithm is very efficient, since most transmissions are unicast and the algorithm tries to advance using the shorter route toward the destination, it has a main drawback: its reliability fully depends on all forwarding nodes choosing a viable next hop node. Because, if a wrong next hop is selected at any step of the path, the forwarded packet is lost.

\begin{figure}[t!]
    \centering
    \includegraphics[width=0.6\textwidth]{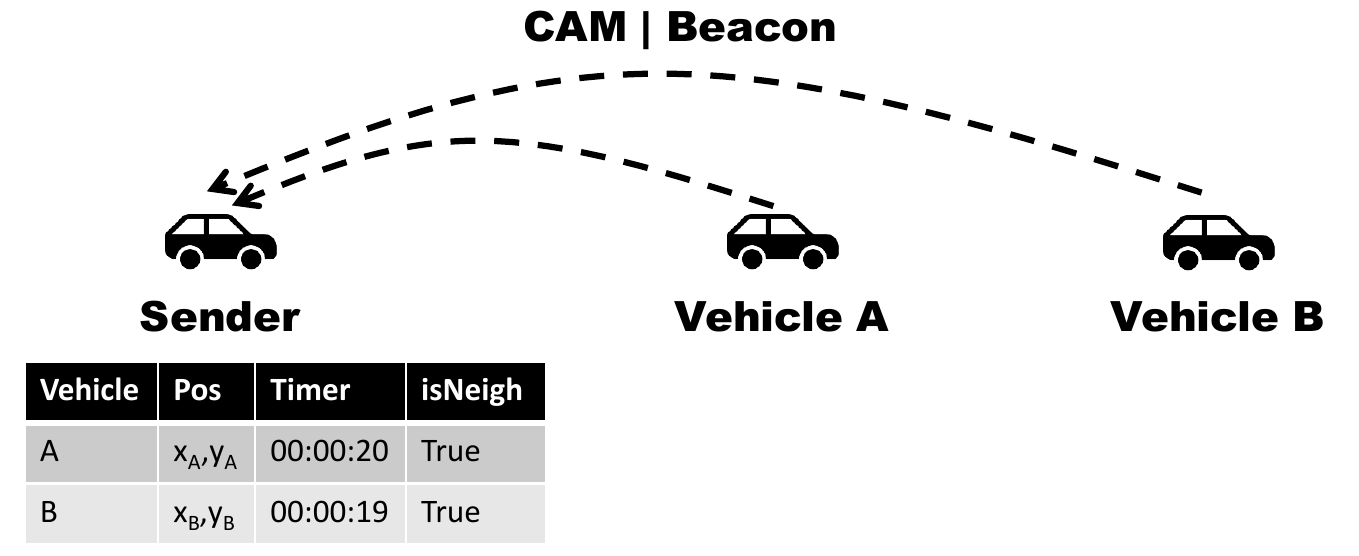}
    \caption{Learning neighbor positions through CAM or Beacon messages}
    \label{fig:Greedy_CAM_Beacon}
\end{figure}

Therefore, it is vital that the sender node is able to select an appropriate one-hop neighbor as the next hop for the packet. However, to do so, the only information a forwarding node has is its \acf{LocT}, where it stores the position vectors of its neighbors. The information about neighbor vehicles is usually obtained by receiving CAMs or Beacon messages. {\color{black}The Beacon service is specified in \cite{etsiNewGeoNetworking}, which defines that Beacon messages are sent as one hop broadcast messages to advertise each node coordinates} (as represented in Fig.~\ref{fig:Greedy_CAM_Beacon}). Beacon messages must be sent periodically (every 3\,s) when using Greedy Forwarding unless a CAM message has been sent less than 1 second ago.    

\begin{figure}[b!]
    \centering
    \includegraphics[width=0.8\textwidth]{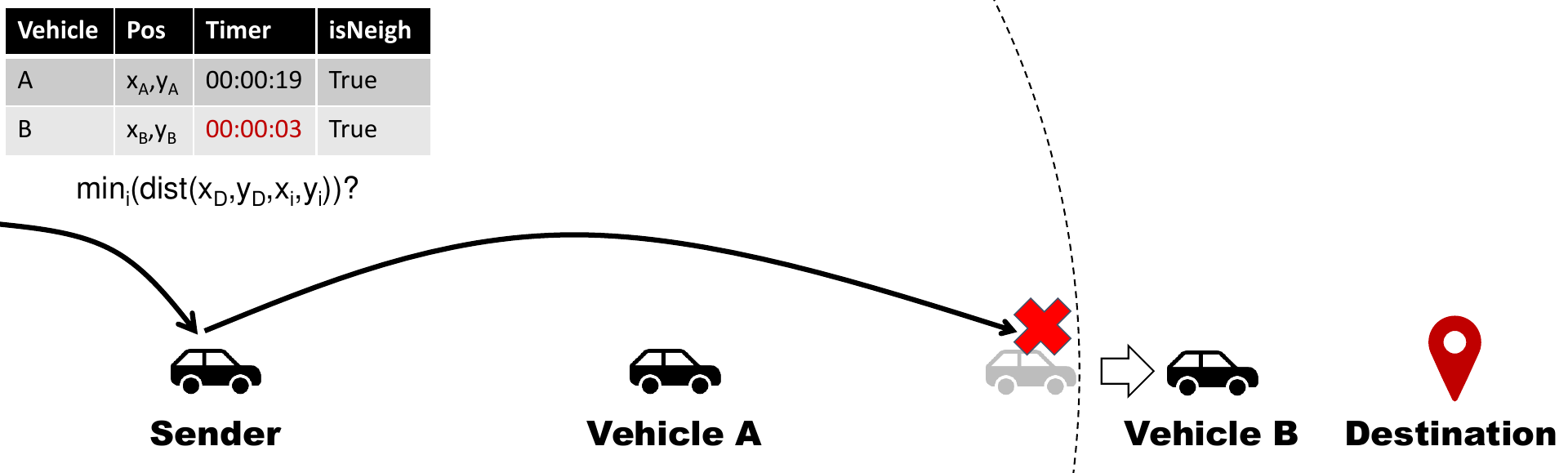}
    \caption{Stale neighbor in Location Table}
    \label{fig:Greedy_Neigh_Flag}
\end{figure}

After receiving a CAM or a Beacon message, the information about the neighbor node, including its coordinates, is stored in the \ac{LocT}. The \ac{LocT} contains the neighbor's position vector and flags that indicate the accuracy of the information and whether a node is actually a neighbor ($is\_neighbor$ flag). If an entry of a neighbor is not updated by a CAM or Beacon, it is dropped after 20 seconds~\cite{etsiNewGeoNetworking}. This means that the information about the coordinates of neighbor nodes may be stale by the time a node tries to forward a packet using Greedy Forwarding. For instance, imagine a vehicle that goes in our same direction, but faster than us (e.g., Vehicle B in Fig.~\ref{fig:Greedy_Neigh_Flag}). While it is inside our reception range, we can hear its CAMs, but once it leaves it, we will stop receiving updates about its position, keeping its last known position in our \ac{LocT}, probably on the edge of our reception range. 

Thus, if several seconds later (up to 20 seconds since the last received Beacon or CAM) we receive another packet to be forwarded using Greedy Forwarding with a destination in the same direction as us, we may choose that farthest vehicle as the best next hop, but since it is no longer inside our transmission range, it could not be reached, regardless of how many unicast re-transmissions we tried, and the packet is finally lost. 

The work in~\cite{riebl2021} identified this problem, and proposed setting the $is\_neighbor$ flag to $false$ every 3.75\,s (the time between two GeoNetworking beacons plus the maximum jitter according to Annex H in~\cite{etsiNewGeoNetworking}). This way, the Greedy Forwarding algorithm can work with a more reliable neighbor flag, thus reducing the probability that the position of that node is stale.

\subsubsection{Too greedy neighbor selection by Greedy Forwarding}

Other ETSI-related works in the literature \cite{Kuhlmorgen:2015, Sandonis2016} have identified the shortcomings of choosing as next hop the neighbor with the maximum progress toward the destination. In high-density scenarios, trying to maximize the progress causes forwarding nodes to choose as next hops vehicles that are on the very border of its transmission range. Thus, even with multiple retransmissions, it may happen that the packet cannot be delivered and is finally lost.  

\begin{figure}[h!]
    \centering
    \includegraphics[width=0.8\textwidth]{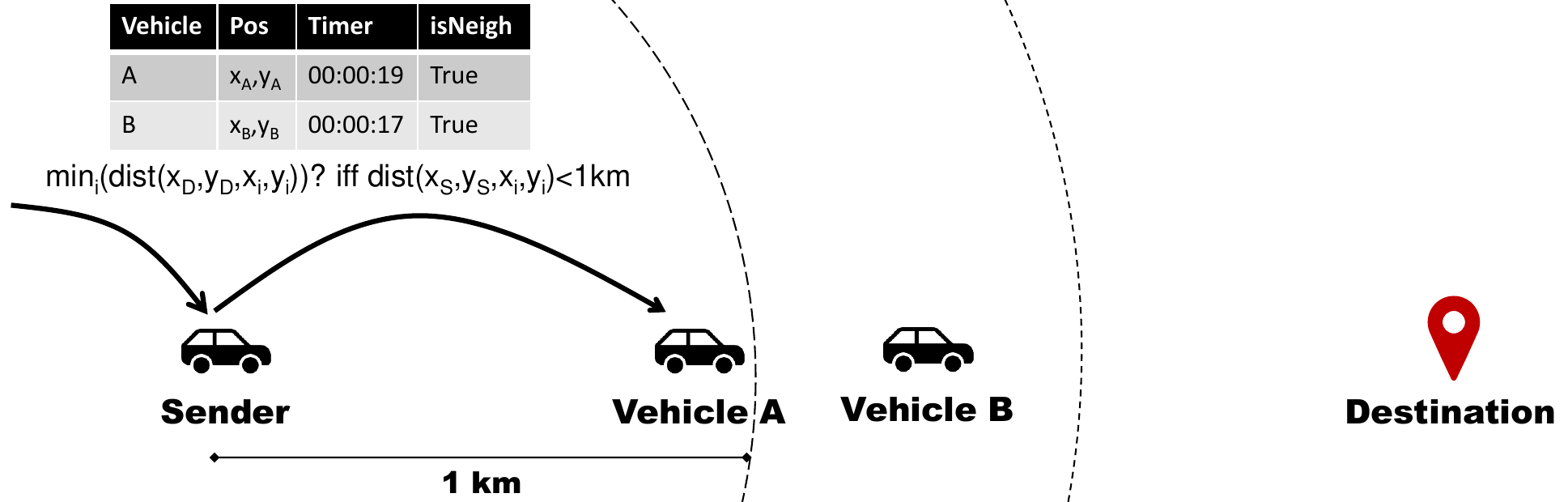}
    \caption{Choosing closer neighbors as next hops}
    \label{fig:Greedy_Rings}
\end{figure}

The proposed solution is quite simple: just define an upper limit for the distance (e.g., 1\,km as illustrated in Fig.~\ref{fig:Greedy_Rings}) at which the next selected hop can be located. Therefore, the packet may perform more hops but it is less probable that the selected next hop is not reachable.

\subsection{ETSI Non-Area \acl{CBF}}

\begin{figure}[tbh!]
    \centering
    \includegraphics[width=0.8\textwidth]{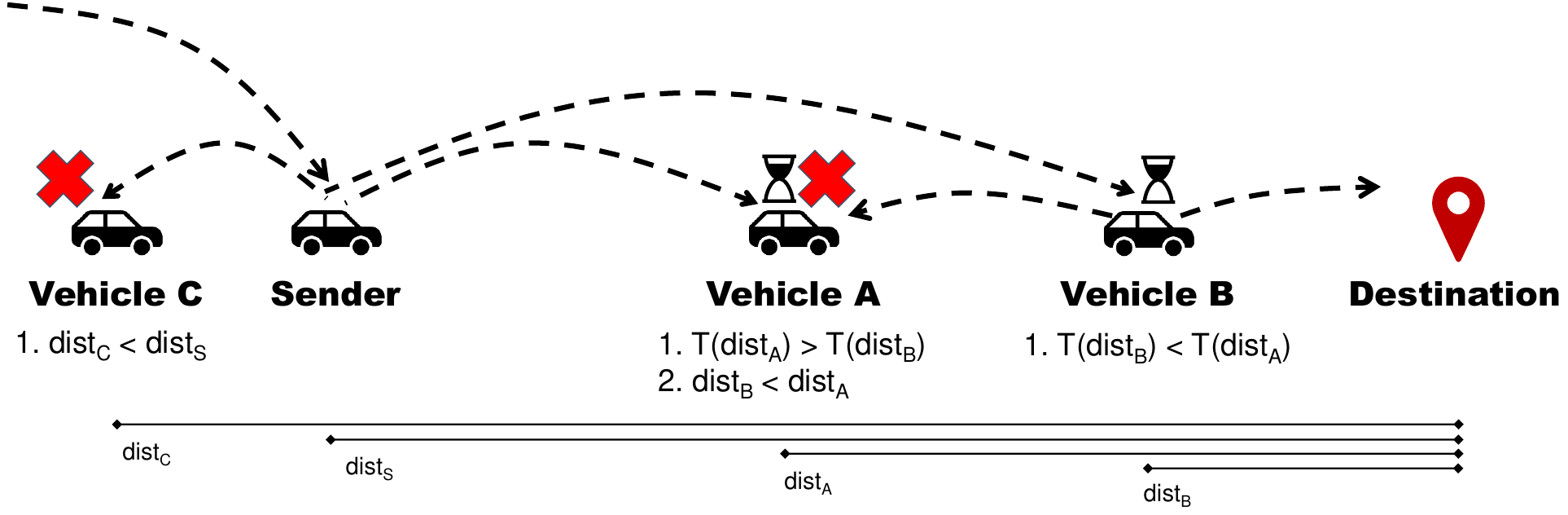}
    \caption{Non-Area Contention-Based Forwarding (CBF)}
    \label{fig:Non-Area_CBF}
\end{figure}

The other forwarding algorithm defined by the ETSI ITS architecture to forward packets toward a Destination Area when the sender is not already inside the area is called Non-Area \ac{CBF}, and it is a small variation of the Area \ac{CBF} algorithm employed for disseminating a message inside a Destination Area. Both \ac{CBF} algorithms are receiver-driven. That is, the sender does not choose who is the next hop as the Greedy Forwarding algorithm does, but it just broadcasts the packet (e.g., as Sender vehicle in Fig.~\ref{fig:Non-Area_CBF}) and the receivers of that packet are the ones who decide who should forward it (e.g., Vehicles A, B). They do so by storing the received packet in the so-called CBF buffer, with an associated timer that specifies when it should be transmitted. However, if another vehicle retransmits the packet before this timer expires, it cancels the transmission of this packet at the remaining vehicles, which drop the packet from their CBF buffer (e.g., the Vehicle B transmission cancels packet in Vehicle A CBF packet buffer).

The main difference between Area and Non-Area \ac{CBF} stems from the difference between their respective goals. Non-Area \ac{CBF} aims to deliver a message to a remote Destination Area, and Area \ac{CBF} looks to distribute a message within that area. The \ac{CBF} timer calculation follows essentially the same logic to obtain shorter timeouts for more distant forwarding candidates. The longer the distance, the smaller the timer duration (i.e., 1\,ms for nodes 1\,km away, up to 100\,ms for vehicles beside the sender), in order to make the nodes further away from the sender to transmit first and cancel the rest of potential forwarders, thus reducing the amount of transmission to cover the whole area.
For Area CBF, the distance that determines the value of \ac{CBF} timeout is the one between the candidate and the last forwarder. In Non-Area CBF the timer duration depends on the progress toward the center of the Destination Area (e.g., $T(dist)$ in Fig.~\ref{fig:Non-Area_CBF}). Nodes closer to the destination have a smaller timeout than the ones near the sender. Besides, all vehicles that are further away from the area center than the previous sender just ignore the packet and do not store it in their CBF buffers (e.g., Vehicle C in Fig.~\ref{fig:Non-Area_CBF}).

This receiver-driven Non-Area CBF algorithm is more robust than the Greedy Forwarding one, in the sense that a forwarding node does not have to know and choose a viable next hop, but it only has to broadcast the packet and all suitable nodes that receive it are able to forward it. On the other hand, Non-Area CBF is much less efficient than Greedy Forwarding based on the number of transmissions because the canceling mechanism does not always work, and it is common for several nodes to forward the same packet toward the Destination Area. The use of timers to trigger each packet forwarding also means that the end-to-end delay tends to be larger in Non-Area CBF than in Greedy Forwarding.

\subsubsection{S-FoT+}

The ETSI Area CBF algorithm has been evaluated in recent works \cite{Amador2022,S-FoT+:2023,Paulin2015}. The work in~\cite{Amador2022,S-FoT+:2023} identifies several phenomena affecting CBF due to the interaction between layers in the architecture that make ETSI CBF inefficient, and proposes improvements to solve the detected problems. In particular, the performance of ETSI Area CBF is affected by the interaction between the Network \& Transport (where GeoNetworking resides) and Access layers. The difference between the time a packet is expected to be transmitted and the actual time it occurs --- since a transmission can be delayed by DCC --- causes different phenomena explored in~\cite{Amador2022,S-FoT+:2023,Paulin2015}, which in turn affects the behavior of CBF, drifting away from what it is expected toward a highly inefficient performance. For example, the optimal forwarder might see its transmission delayed by DCC, and then sub-optimal forwarders do not cancel their CBF timers, send their packets to the Access layer and, given a set of circumstances, forward a message that breaks the expected system behavior.

The first phenomenon is the existence of an excessive number of duplicate packets on the medium. The ETSI specification explicitly offloads \ac{DPD} tasks --- which are performed before starting the execution of other mechanisms --- to the CBF algorithm. The idea is that an optimal forwarding can inhibit sub-optimal candidates that still have a copy of the message in their \ac{CBF} buffers. If a received packet is not in the buffer, it is assumed that it is a new packet. The ETSI CBF mechanism does not keep a record of past messages, and it expects the most optimal forwarder to inhibit all other candidates, thus not requiring such a record. However, since the Access layer might delay a packet transmission, a copy of the message might arrive at a node which has its own copy waiting to be transmitted at the Access layer or being already sent (thus, not in the CBF buffer), and will be buffered again. The first copy will be transmitted eventually, and there is a chance for the duplicate to be forwarded again if it is not canceled in the CBF buffer. Then, a node can unknowingly forward the same message several times. The works in~\cite{Amador2022,Paulin2015,riebl2021} describe this phenomenon and propose the inclusion of a \ac{DPD} mechanism. 

The works in~\cite{Amador2022,Paulin2015}  also identify a second phenomenon: sub-optimal forwarders can inhibit better candidates from forwarding a message. This happens when a sub-optimal forwarder gets access to the medium before an optimal candidate can forward a message. The work in~\cite{Paulin2015} proposes the use of probabilities to decide if a message gets its CBF timer canceled if a copy is received. The proposal in~\cite{Amador2022} is to determine if the copy comes from a better forwarder with complete certainty, with a mechanism called \ac{GPC}. If so, the message is canceled. Otherwise, if the message comes from a sub-optimal forwarder or an unknown sender, the CBF timer is updated but not canceled, allowing the optimal forwarder to help the message advance. 

A different phenomenon identified in~\cite{Amador2022} is the potential for losses in the first hop. If a source node had its original message collide, there are no options for recovery. Thus, \cite{Amador2022} proposes for the source node to buffer its own packet. In case it does not hear from a forwarder after the maximum possible CBF timeout, it transmits the packet once again.

Furthermore, the work in~\cite{Amador2022} proposes that the CBF timers become aware of the status of the Access layer, specifically, of the time to the next transmission allowed by DCC. The literature identifies a problem where GeoNetworking sends a packet down to the Access layer and, if DCC is restricting transmissions, it has to wait to be transmitted. In some cases, the message might not even be transmitted, since forwarding occurs with low priority. \ac{FoT}, presented in~\cite{Amador2022} uses the Cross-layer DCC information available at the Management Entity to: 1) calculate the CBF timer (choosing the maximum between the distance-based timer and the time to the next allowed transmission), and 2) decide when to move a packet from the CBF buffer to the Access layer (i.e., only if the time to the next allowed transmission is zero, or if it has already passed and no transmission occurred, as proposed in the FoT+ mechanism from~\cite{S-FoT+:2023}). Thus, if a candidate with better possibilities to access the medium is able to forward the packet (thus becoming the best candidate when network congestion is high), the rest of the candidates, if they are worse forwarders, will be able to cancel their copies in their CBF buffers. 

\ac{LocT} management issues may also affect \ac{CBF}. The work in~\cite{Amador2022} identifies the impact of the combination of Area \ac{CBF} with Greedy Forwarding. The GeoNetworking specification asks to discard messages received by nodes outside of the Destination Area from senders within the area. However, due to stale information in the \ac{LocT}, a receiver may consider an invalid message as valid (i.e., because it believes a sender is outside of the area when it is really inside). These CBF messages are addressed to the broadcast address, and Greedy Forwarding usually sends them to unicast addresses. Furthermore, several nodes outside the area receive the packet, and try to forward it immediately, causing collisions. The work in~\cite{Amador2022} proposes a mechanism in which broadcast packets received outside the Destination Area are never forwarded. However, this Unicast inhibition can affect the Greedy Forwarding algorithm, which uses broadcast in cases where it cannot find a better next hop (e.g., when the forwarder is a local maximum).

\begin{figure}
    \centering
    \includegraphics[width=0.8\linewidth]{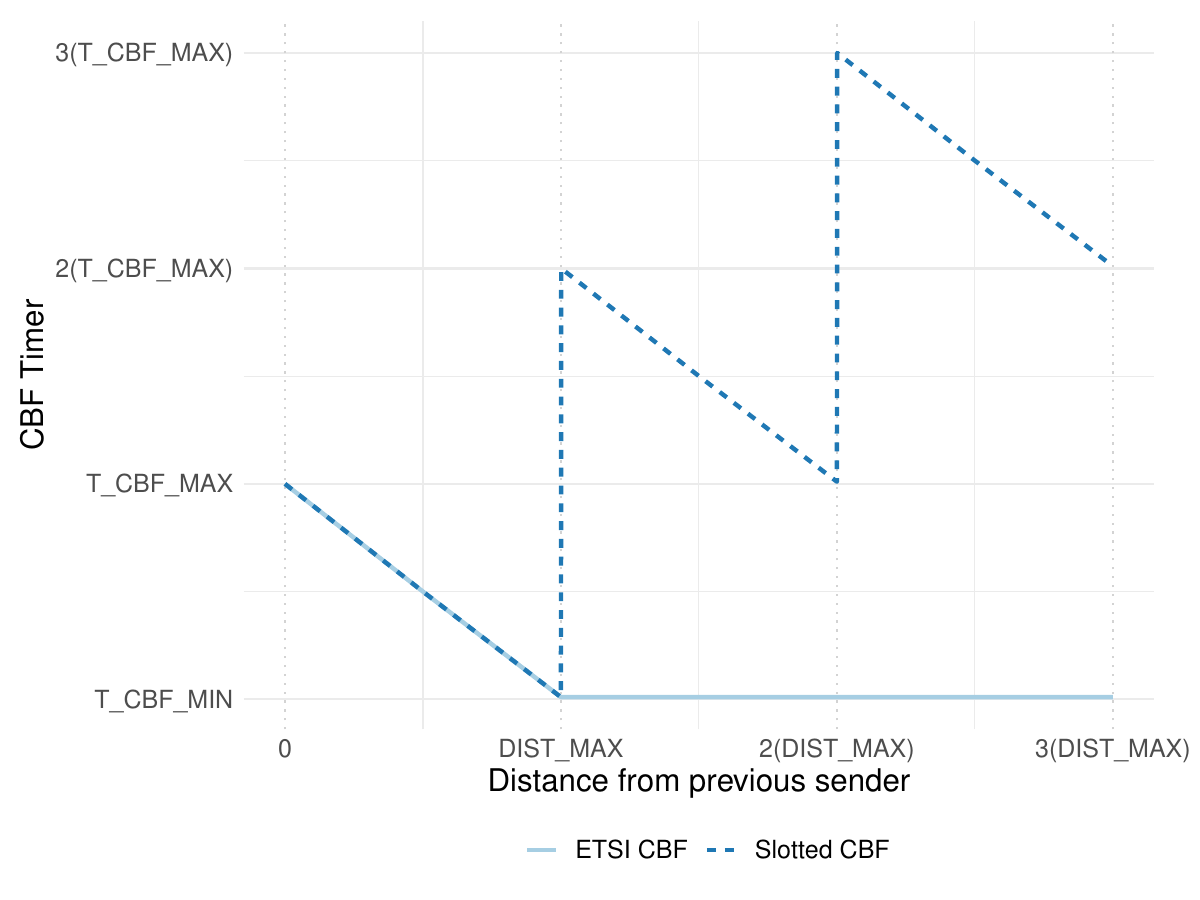}
    \caption{CBF timer pattern for ETSI CBF and Slotted CBF}
    \label{fig:s_cbf_pattern}
\end{figure}

Finally, a phenomenon that can occur due to the expected communication ranges being exceeded is explored in~\cite{S-FoT+:2023}. If there are nodes that receive a packet beyond the maximum expected communication range ($DIST_{MAX} = 1000m$), they will all decide to buffer the packet for the minimum time possible, and probably transmit at the same time, causing a collision. Authors propose a timeout calculation process called \ac{S-CBF}, where the distance from the sender is divided into \textit{slots} of size $DIST_{MAX}$ starting from zero. Each slot number becomes a multiplying factor for the number of times $T_{CBF-MAX}=100ms$ is added to the timeout calculated in a modified CBF formula (\textit{Slot 0} timeouts range from 1--100\,ms, \textit{Slot 1} timeouts go between 101--200\,ms, etc.)\textcolor{black}{, as illustrated by the patterns in Fig.~\ref{fig:s_cbf_pattern}}. Thus, avoiding the synchronization of transmissions from forwarders beyond $DIST_{MAX}$. The resulting mechanism, that includes \ac{DPD}, \ac{GPC}, S-CBF, Unicast inhibition, and DCC awareness (FoT+) is called S-FoT+~\cite{S-FoT+:2023}.

\section{Evaluation Parameters} %% Óscar
\label{sec:evaluation}

\subsection{Combination of Non-Area and Area Mechanisms}
\label{subsec:combinations}

We evaluate different combinations of non-area and area forwarding algorithms. Table~\ref{tbl:setups} shows the set-ups we evaluate and the names by which we will refer them in the remainder of this work. Table~\ref{tbl:setups} also shows if certain features are present in a given forwarding algorithm, such as \ac{DPD}, geographical awareness (e.g., \ac{GPC}), \ac{DCC} awareness (e.g., \ac{FoT}+), or if there are limits to how far a hop is expected to reach (e.g., \ac{S-CBF})

\begin{table*}[tbh!]
\centering
\small
\begin{tabularx}{\textwidth}{| >{\raggedright\arraybackslash}X ||  >{\raggedleft\arraybackslash}X | >{\raggedleft\arraybackslash}X |
>{\raggedleft\arraybackslash}X | >{\raggedleft\arraybackslash}X | >{\raggedleft\arraybackslash}X | >{\raggedleft\arraybackslash}X | >{\raggedleft\arraybackslash}X | >{\raggedleft\arraybackslash}X | >{\raggedleft\arraybackslash}X | >{\raggedleft\arraybackslash}X | }
\hline
\multirow{2}{*}{\textbf{Set-up}}& \multicolumn{6}{c|}{\textbf{Non-Area}} & \multicolumn{4}{c|}{\textbf{Area}} \\\cline{2-11} & \textbf{Fwd. Alg.} & \textbf{DPD} & \textbf{Prog. Check} & \textbf{Range limit} & \textbf{Unicast inhibition} & \textbf{DCC awareness} & \textbf{Fwd. Alg.} & \textbf{DPD} & \textbf{Prog. Check} & \textbf{DCC awareness}  \\\hline
ETSI CBF & Non-Area CBF & No  & When buffering & No & NA & No & Area CBF  & No & No & No\\ \hline
Greedy and CBF & ETSI Greedy Forwarding & Yes & NA & No & No & No & ETSI Area CBF & No & No & No\\ \hline
\mbox{S-FoT+} & Non-Area \mbox{S-FoT+} & Yes & When buffering and canceling  & Yes\dag & NA & Yes & Area \mbox{S-FoT+} & Yes & Yes & Yes \\ \hline
Greedy+ and \mbox{S-FoT+} & Greedy+ & Yes & NA & Yes\ddag & Yes & No & Area \mbox{S-FoT+} & Yes & Yes & Yes\\ \hline
\multicolumn{11}{l}{\dag S-FoT+ uses S-CBF~\cite{S-FoT+:2023}, where $DIST_{MAX}$ is a multiplying factor for CBF timer calculation.}\\
\multicolumn{11}{l}{\ddag Greedy+ chooses the farthest valid neighbor within $DIST_{MAX}$ as the next hop}
\end{tabularx}
\caption{Characteristics of the evaluated set-ups}
\label{tbl:setups}
\end{table*}

Greedy+ is a modification of the Greedy Forwarding algorithm defined in the ETSI GeoNetworking specification~\cite{etsiNewGeoNetworking}. It includes two optimizations:
\begin{enumerate}
    \item A variation of the correction proposed in~\cite{riebl2021}, where the $is\_neigbor$ flag is set to $false$ after 1000\,ms (similar to the logic~\cite{riebl2021} followed with the Beacons, but here we use the maximum time between two consecutive \acp{CAM}).
    \item The range limitation proposed in~\cite{Sandonis2016}, and we set the farthest distance the next hop can be from the sender to $DIST_{MAX}$ (i.e., 1000\,m for our experiment).
\end{enumerate}

S-FoT+ is specified in \cite{S-FoT+:2023} as an area forwarding algorithm. In these experiments, we use it also for non-area forwarding. The only adaptation is the calculation of the distance used to determine the CBF timer, which in area forwarding is based on the advance from the previous sender, while in non-area forwarding is the progress toward the Destination Area (the difference between the distance from the previous sender to the destination, and the distance from the candidate forwarder to the destination). Following the philosophy of \ac{GPC}, the progress check is used for both buffering (a packet is buffered only if the candidate vehicle represents progress compared with the previous sender), and packet cancellation (only better forwarders can cancel a packet, otherwise the CBF timer is updated).

\subsection{Simulation Parameters}
\label{subsec:simpars}

\begin{table}[h]
	\centering
	\begin{tabular}{| l | l |}
		\hline
		\textbf{Parameter}  & \textbf{Values} \\
		\hline
		Access Layer protocol & ITS-G5 (IEEE 802.11p) \\
		Channel bandwidth & 10\,MHz at 5.9\,GHz \\
		Data rate & 6\,Mbit/s \\
            DCC & ETSI Adaptive DCC \\
		Transmit power & 20\,mW \\
		Path loss model & Two-Ray interference model\textcolor{black}{~\cite{Sommer:2012}} \\
		Maximum transmission range measured & 1500\,m \\
            Unicast Transmission Retry Limit & 7 \\
		CAM packet size & \textcolor{black}{ \{85,285\} bytes (ETSI CAM~\cite{etsiCA} as implemented in~\cite{Artery})} \\
            \textcolor{black}{CAM generation frequency} & \textcolor{black}{Variable rate 1--10~Hz (ETSI CAM~\cite{etsiCA})} \\
		CAM Traffic Class & TC2 \\
		DENM packet size & 301 bytes \\
		DENM Traffic Class & TC0 (Source) and TC3 (Forwarders) \\
		DENM lifetime & 10\,s\\
		DPL size & 32 packet identifiers per Source\\
            Default Hop Limit & 10\\
		\hline
		\multicolumn{2}{| c |}{\textit{Urban Scenario}}\\\hline
		Obstacle model & Simple obstacle shadowing\\
		Obstacle parameters & Buildings: 9\,dB/cut, 0.4\,dB/m\\\hline
	\end{tabular}
 	\caption{Simulation Parameters}
	\label{tbl:simpars}
\end{table}

We evaluate the different combinations from Section~\ref{subsec:combinations} using Artery~\cite{Artery}. Artery runs on OMNET++, and implements the ETSI-ITS protocol stack through its Vanetza module\footnote{\textcolor{black}{The implementation for Greedy+ and S-FoT+ are available in \href{https://github.com/oscarmex1986/vanetza/tree/S-FoT\%2B}{https://github.com/oscarmex1986/vanetza/tree/S-FoT\%2B}}}. The physical model we use is based on Veins~\cite{Veins}. Vehicles are controlled by the microscopic mobility simulator SUMO~\cite{sumo2012}. Table~\ref{tbl:simpars} shows the simulation parameters used on both the urban and the highway scenarios. All simulations are repeated 5 times with different seeds and the results show the average and 95\% confidence interval of the 5 simulation results.

The highway scenario consists of a 5\,km road with 5 lanes on each direction. The source vehicle is located on the left end of the road, and generates \acp{DENM} at a 1\,Hz rate after a warm-up period of at least 150\,s. The source is outside of the Destination Area, which is a rectangle covering 2\,km on both lane directions starting 2.5\,km away from the source node (i.e., the message has to travel 2.5\,km outside of the area and then cover 2\,km inside the Destination Area). Measurements are taken for 30\,s and five different vehicle densities are tested (10--50\,veh/km per lane).

For the urban scenario, we use a map of central Madrid, sourced from OpenStreetMap~\cite{OSM}. The scenario consists of small streets, medium-sized avenues, and long, multi-lane roadways. \textcolor{black}{There are 1800 vehicles in the scenario}. A source node is located on the southwest of the map and two experiments, with two different Destinations Areas, are performed. We use the same data set\footnote{\textcolor{black}{The SUMO scenario (map and vehicle traffic) is available in \href{https://github.com/igsoto/sumo-map-madrid}{https://github.com/igsoto/sumo-map-madrid}}.} as in~\cite{Uruena2017}, where each vehicle is assigned a trip with a minimum length of 1\,km. Trips follow routes calculated by the Dynamic User Assignment (DUA) \cite{Gawron1999} and A-star \cite{a-star} algorithms, which distribute trips in a way that does not send all vehicles through main arteries disregarding smaller streets.

Finally, all vehicles in the scenario run the \ac{CA} and \ac{DEN} basic services. Vehicles send \acp{CAM} following the kinematic rules from~\cite{etsiCA}. Any node can potentially become a forwarder for a multi-hop \ac{DENM}. Additional characteristics for the highway and urban scenarios are explained in Section~\ref{sec:highway} and ~\ref{sec:urban}, respectively.

\subsection{Performance Metrics}
\label{subsec:metrics}

We evaluate the performance of the protocol combinations defined in Table~\ref{tbl:setups} by measuring efficacy (e.g., whether messages are delivered successfully to nodes in the Destination Area) and efficiency (e.g., how many transmissions are used). The specific metrics are described below:
\begin{itemize}
    \item \textbf{\ac{PDR}:} the number of vehicles in the Destination Area receiving a message successfully divided by the number of vehicles in the area at the instant the message was generated.
    \item \textbf{End-to-end (E2E) delay:} the time difference between the timestamp of a message reception by a node in the Destination Area and the time it was generated at the source node. Both timestamps are set at the Facilities layer. Where indicated, E2E delay is measured for the first reception within the Destination Area.
    \item \textbf{Number of transmissions:} the combined number of frame transmissions required for a single message to attempt reaching (non-area forwarding) and covering the Destination Area (area forwarding).
\end{itemize}

\section{Evaluation of Highway Scenarios} 
\label{sec:highway}

\begin{figure}[tbh!]
    \centering
    \includegraphics[width=0.9\textwidth]{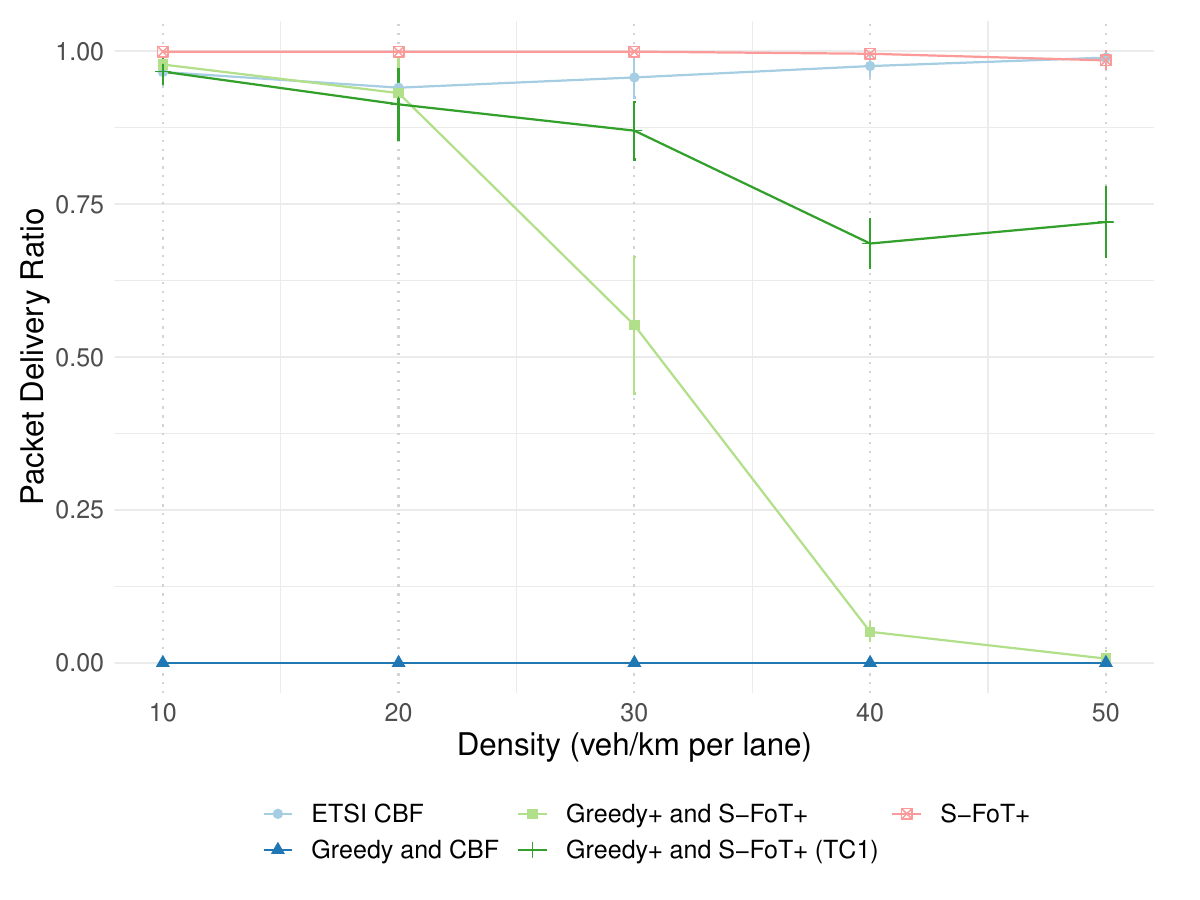}
    \caption{Average Packet-delivery Ratio in the Destination Area}
    \label{fig:pdr_hwy}
\end{figure}

Fig.~\ref{fig:pdr_hwy} shows the results for \ac{PDR} measurements within the Destination Area. There is one main takeaway from this result: Greedy Forwarding (see Greedy and CBF combination), as specified in~\cite{etsiNewGeoNetworking}, fails to even reach the Destination Area. Even with the help of \ac{ARQ} at the MAC sublayer, reliability is null. This is due to the default 20\,s timer of \ac{LocT} entries, which increases the probability of the node selecting a "stale" neighbor. This acts in combination with the "greedy" nature of ETSI Greedy Forwarding, which causes the algorithm to select the farthest neighbor on the \ac{LocT}. Thus, even a "fresh" neighbor (which has just sent a CAM) is likely far enough to cause reliability issues. It must be also noted that this result is obtained with homogeneous vehicle radios, where the sense of "neighborhood" is symmetric, but even with this favorable assumption, ETSI Greedy Forwarding fails completely to reach remote Destinations Areas. In practice, differences between transmission equipment and antennas might mean that a vehicle perceived as a neighbor in reception cannot be reached in transmission, leading to even more lost packets. 

The rest of the figure can be analyzed in two parts: 1) low densities 10--20, where \ac{DCC} is not a factor, and 2) high densities 30--50, where \ac{DCC} starts restricting the rate at which nodes can send messages. For the first two densities, Greedy+ (in the Greedy+ and S-FoT+ combination) exhibits the results of keeping a better track of neighbors and selecting only the farthest possible within $DIST_{MAX} = 1\,km$. However, there is an interesting phenomenon, reflected on the same densities in Table~\ref{table:txd_hwy}, which shows the average number of transmissions per message of the different algorithms: Greedy+ combined with S-FoT+ has more transmissions --- and worse \ac{PDR} --- than pure S-FoT+. This can be attributed to the absence of the Unicast inhibition feature: packets, addressed to the Broadcast MAC address, that are received from vehicles erroneously thought to be outside of the area (e.g., the last CAM received indicated that the vehicle was outside, but it is now inside the Destination Area) are then forwarded by nodes outside the area as unicast messages toward the area, causing collisions as described in~\cite{Amador2022}.

\begin{table*}[tbh!]
\centering
\begin{tabularx}{\textwidth}{| >{\raggedright\arraybackslash}X |  >{\raggedleft\arraybackslash}X |
>{\raggedleft\arraybackslash}X | >{\raggedleft\arraybackslash}X | >{\raggedleft\arraybackslash}X | >{\raggedleft\arraybackslash}X |}
\hline
\textbf{Density (veh/km per lane)} & \textbf{ETSI CBF} & \textbf{Greedy and CBF} & \textbf{S-FoT+} & \textbf{Greedy+ and S-FoT+} &\textbf{Greedy+ and S-FoT+ (TC1)} \\\hline
10 & 767.97  & 8.33  & 25.57  &  45.88  & \textcolor{black}{54.76} \\ \hline
20 & 885.86  & 8.15  & 53.64  & 143.27  & \textcolor{black}{216.62} \\ \hline
30 & 845.72  & 8.15  & 69.63  & 118.83  & 262.15       \\ \hline
40 & 826.62  & 7.96  & 51.32  &   7.87  &  65.50       \\ \hline
50 & 864.54  & 7.75  & 63.15  &   5.49  &  98.24       \\ \hline
\end{tabularx}
\caption{Average number of transmissions per message in the highway scenario (area and non-area)}
\label{table:txd_hwy}
\end{table*}

\begin{table*}[tbh!]
\centering
\begin{tabularx}{\textwidth}{| >{\raggedright\arraybackslash}X |  >{\raggedleft\arraybackslash}X |
>{\raggedleft\arraybackslash}X | >{\raggedleft\arraybackslash}X | >{\raggedleft\arraybackslash}X | >{\raggedleft\arraybackslash}X |}
\hline
\textbf{Density (veh/km per lane)} & \textbf{ETSI CBF} & \textbf{Greedy and CBF} & \textbf{S-FoT+} & \textbf{Greedy+ and S-FoT+} &\textbf{Greedy+ and S-FoT+ (TC1)} \\\hline
10 & 0.0629  & ---  & 0.0745  & 0.0094  & \textcolor{black}{0.0035} \\ \hline
20 & 0.1151  & ---  & 0.0795  & 0.0229  & \textcolor{black}{0.0047} \\ \hline
30 & 0.2164  & ---  & 0.0827  & 0.5837  & 0.1007 \\ \hline
40 & 0.2040  & ---  & 0.1574  & 3.2390  & 0.4376 \\ \hline
50 & 0.1949  & ---  & 0.2213  & 6.5602  & 0.5657 \\ \hline
\end{tabularx}
\caption{Average End-to-End delay (in seconds) for the first successful packet arrival into the Destination Area (highway)}
\label{table:e2e_hwy}
\end{table*}

Starting on the 30~veh/km per lane density (D30), Greedy+ starts exhibiting a degradation in performance. Its \ac{PDR} falls significantly on each density increase, which can have two possible explanations: 1) worse transmission reliability due to increased traffic, and 2) the effect of \ac{DCC}. Table~\ref{table:e2e_hwy} shows the time it takes for a message to arrive from the source to the first node they reach within the Destination Area. At D30, the latency starts to increase significantly. When analyzing latency from D30 onward, in combination with the same region of Table~\ref{table:txd_hwy}, it can be inferred that \ac{DCC} prevents packets from being relayed, due to the fact that forwarding packets have lower priority in the \ac{DCC} queues (TC3 --- the lowest priority) than CAM messages (TC2). This phenomenon is exclusive of Greedy+, since only one node is selected as a next hop, while multiple nodes are potential forwarders in CBF-based mechanisms, so we just need an opportunity to transmit in any of them to continue progressing toward the destination.

\textcolor{black}{To assess the effect of \ac{DCC} on Greedy+, we have performed an additional set of experiments with Greedy+ combined with S-FoT+ with forwarding using TC1 (higher priority than \ac{CAM} with TC2). The results in terms of \ac{PDR} (see Fig.~\ref{fig:pdr_hwy}) improve in medium and high densities in comparison to the same combination using TC3. However, reliability does not reach the level of ETSI CBF and S-FoT+ in densities 20--50. The effect of higher priority on delays (see Table~\ref{table:e2e_hwy})is even noticeable in the lower densities, where Greedy+ TC1 reaches the area in a fraction of the time of its TC3 counterpart. However, for medium and high densities, even if delays are lowered compared to Greedy+ TC3, they are nevertheless above those of ETSI CBF and S-FoT+.} We conclude that this is a consequence of the single-next-hop approach, which only has one forwarder node waiting for \ac{DCC} to allow for a transmission. CBF-based mechanisms have many potential forwarders and the one with the best CBF timeout/\ac{DCC} combination will transmit the message the sooner. While S-FoT+ is aware of the rate allowed by \ac{DCC} and keeps the packets in the CBF buffer enabling its cancellation, ETSI CBF uses brute force to allow for a message to advance, which is reflected in the huge difference in the number of transmissions.

\begin{figure}
    \centering
    \includegraphics[width=1\linewidth]{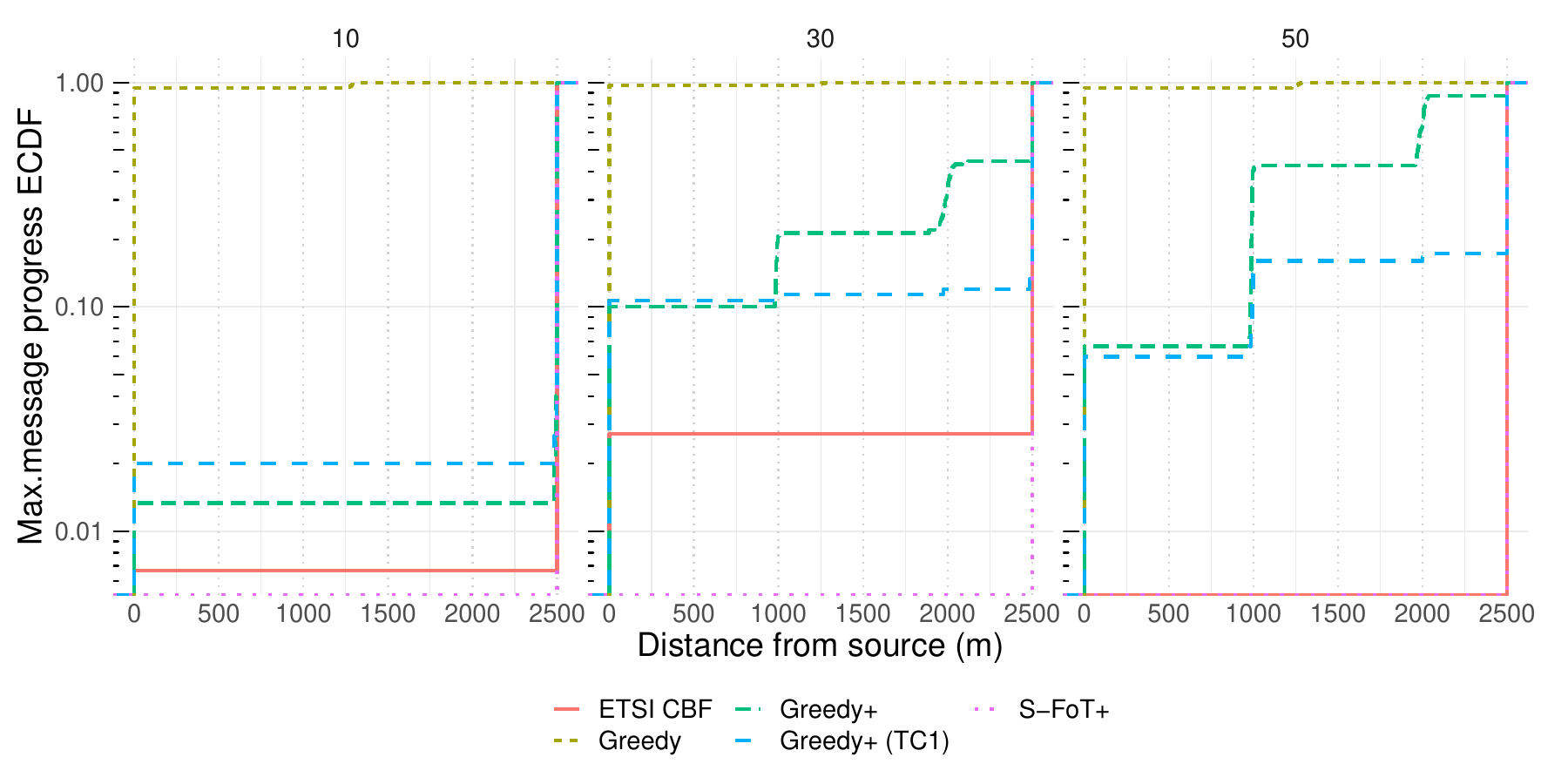}
    \caption{Empirical cumulative distribution function for message losses outside the Destination Area in densities 10, 30, and 50.}
    \label{fig:ecdf_mech_densities}
\end{figure}

\textcolor{black}{To fully understand how non-area forwarding mechanisms perform, we identify where packets that do not reach the Destination Area get stalled. Fig.~\ref{fig:ecdf_mech_densities} shows the empirical cumulative distribution function (ECDF) for messages progress before reaching the Destination Area (located 2500\,m away from the source) for densities 10, 30, and 50. Therefore, Fig.~\ref{fig:ecdf_mech_densities} shows the proportion of packets that reach a certain distance from the source, or less, out of those that do not reach the Destination Area.  It is worth noticing that S-FoT+ never loses a message on the way to the Destination Area, thus, its ECDF stays on zero. The second most reliable mechanism, ETSI CBF, only fails to reach the area when it loses a message because of a collision in the source node --- an issue previously identified in~\cite{Amador2022}. Greedy forwarding fails to reach the Destination Area always, and it is the source node that suffers the bulk of the losses, with only a fraction of the messages completing the first hop at around 1250\,m, where they finally get stalled. This is due to a bad choice for the next hop, and the improvement in performance for Greedy+ indicates so. For D10, where PDR is high, the few losses that occur happen at the source node. However, D30 and D50 show that each hop represents a risk for losses. Greedy+ loses almost half of the packets on the second hop for D30 and on the first for D50. Since we attribute some of these losses to a bottleneck created by DCC, we see this effect with Greedy+ using TC1, where most of the losses occur at the source in D30 --- plus a minimum contribution by subsequent hops---, and are distributed between the source and the first hop for D50. Thus, we show that the shortcomings for Greedy are caused both by bad choices for next hops and the effect of DCC.}

In summary, ETSI CBF and S-FoT+ are the best approaches as mechanisms to reach the Destination Area effectively. However, S-FoT+ also does it efficiently, using only a fraction of the transmissions ETSI CBF uses, with a difference up to an order of magnitude lower (see Table~\ref{table:txd_hwy}). The main takeaway of these results is that, even with optimizations, the trade-off between latency and reliability that is offered by Greedy Forwarding is not such when \ac{DCC} is factored in, thus making contention-based mechanisms both more reliable and, in the case of S-FoT+, more efficient.

\section{Evaluation of Urban Scenarios} 
\label{sec:urban}

\begin{figure}[tbh!]
    \centering
    \includegraphics[width=0.8\textwidth]{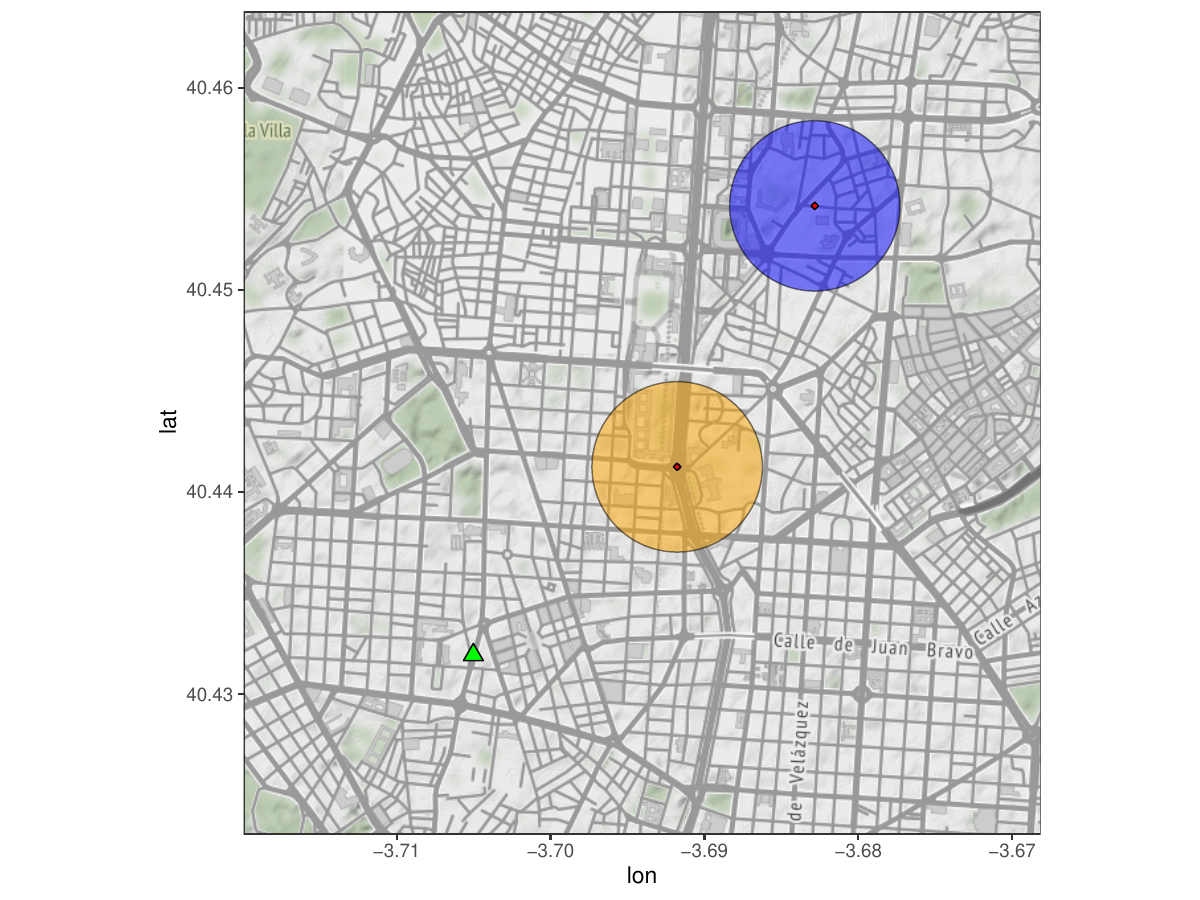}
    \caption{Urban scenario: source (green triangle), and Destination Areas (circles)}
    \label{fig:urban_scenario}
\end{figure}

Fig.~\ref{fig:urban_scenario} shows the map of the urban setting where we evaluate non-area and area GeoNetworking. A vehicle in the southwest region of the map (green triangle) sends \acp{DENM} to a Destination Area. For the first set of measurements, the messages have to reach a circular area of $r=500m$ at the center of the map (yellow circle -- marked as CE in the figures), with its center located approximately 1.5\,km away from the source. For the second set of measurements, the Destination Area is centered approximately 3\,km away, on the northeast region of the map (marked as NE in the figures). We evaluate the same mechanisms as in Section~\ref{sec:highway}. A summary of the results is presented in Table~\ref{tbl:UrbanResults}.

\begin{figure}[tbh!]
    \centering
    \includegraphics[width=0.7\textwidth]{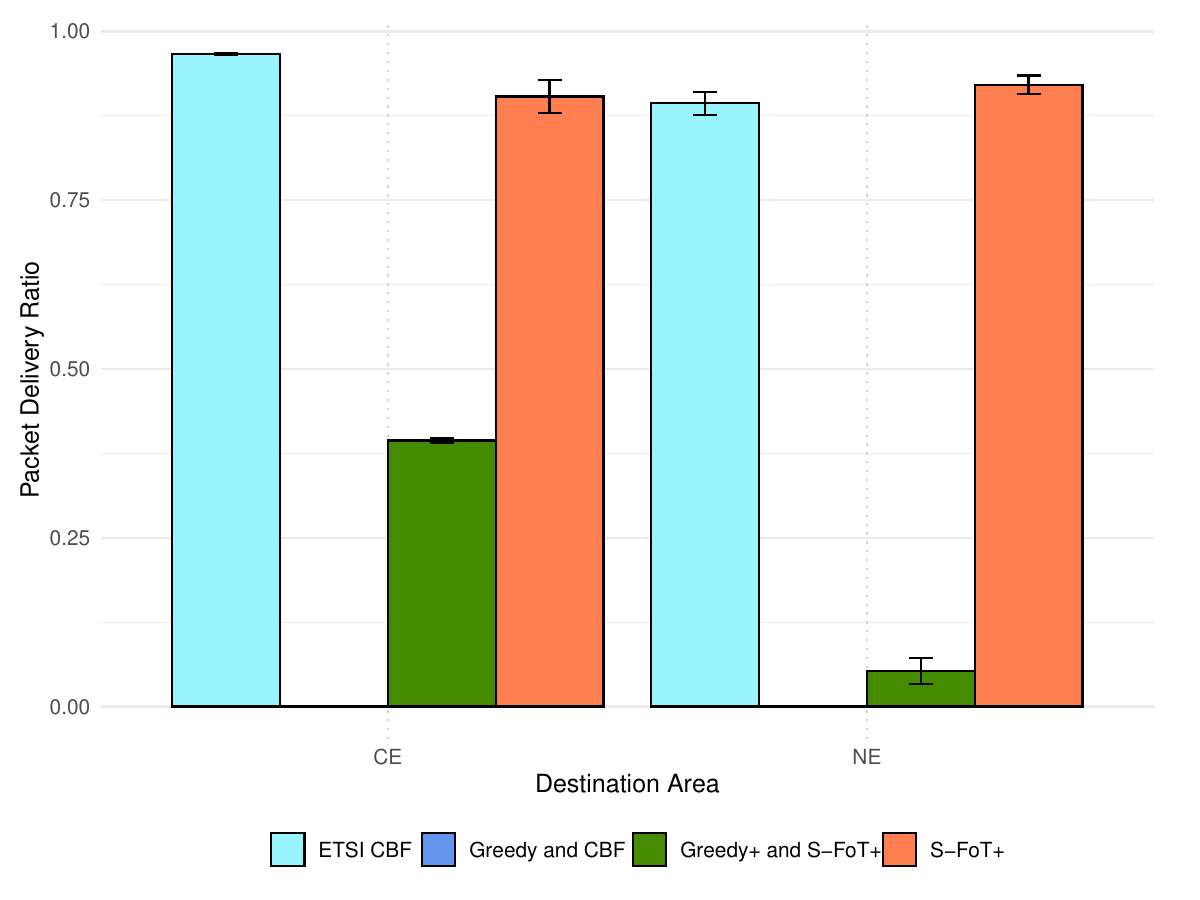}
    \caption{Average Packet-delivery Ratio for the two urban Destination Areas}
    \label{fig:pdr_urban}
\end{figure}

\begin{figure}[tbh!]
    \centering
    \includegraphics[width=0.7\textwidth]{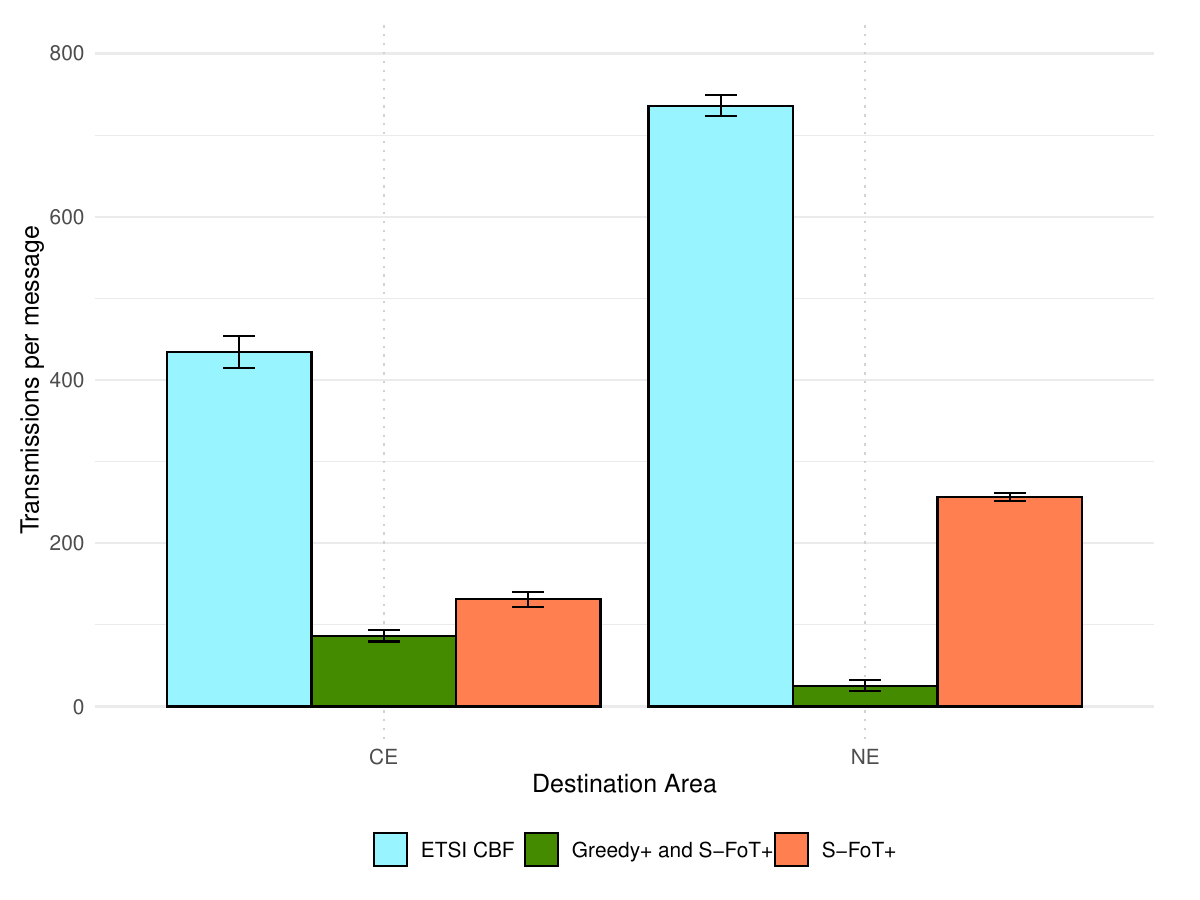}
    \caption{Average number of transmissions per message for two urban Destination Areas}
    \label{fig:txd_da}
\end{figure}

\begin{table*}[tbh!]
\centering
\small
\begin{tabularx}{\textwidth}{| >{\raggedright\arraybackslash}X ||  >{\raggedleft\arraybackslash}X | >{\raggedleft\arraybackslash}X |
>{\raggedleft\arraybackslash}X || >{\raggedleft\arraybackslash}X | >{\raggedleft\arraybackslash}X | >{\raggedleft\arraybackslash}X |  }
\hline
\multirow{2}{*}{\textbf{Set-up}}& \multicolumn{3}{c||}{\textbf{Center (CE)}} & \multicolumn{3}{c|}{\textbf{Northeast (NE)}} \\\cline{2-7} & \textbf{PDR} & \textbf{E2E}\dag & \textbf{Avg. Tx.} & \textbf{PDR} & \textbf{E2E}\dag & \textbf{Avg. Tx}\\\hline
ETSI CBF & 0.9665 & 0.2889\,s & 434.5  & 0.8936 & 0.3798\,s & 735.76  \\ \hline
Greedy and CBF & 0.0000  & --- & 8.2 & 0.0000 & --- & 8.13 \\ \hline
S-FoT+& 0.9035 & 0.2701\,s  & 131.3 & 0.9210 & 0.3610\,s  & 256.94 \\ \hline
Greedy+ and S-FoT+ & 0.3939 & 0.0058\,s & 86.6 & 0.0526 & 0.0102\,s & 25.49 \\ \hline
\multicolumn{7}{l}{\dag End-to-End delay is only measured for messages received by at least one node in the Destination Area} 
\end{tabularx}
\caption{Results for the Urban Scenario}
\label{tbl:UrbanResults}
\end{table*}

Once again, Greedy Forwarding as specified in~\cite{etsiNewGeoNetworking} fails to reach either Destination Area. Fig.~\ref{fig:pdr_urban} shows the average \ac{PDR} for every set-up. Greedy+ also finds hurdles to reach the areas, even if it is able to select "fresh" neighbors within a reasonable distance, and there is a significant drop in success rates for the more distant Destination Area. Fig.~\ref{fig:txd_da} shows the number of transmissions each mechanism uses. \ac{PDR} has to be factored in when interpreting this figure, since Greedy+ combined with S-FoT+ generates fewer transmissions, but the "trade-off" in reliability puts Greedy+ in a different league than purely receiver-based mechanisms. This is reflected in the results for the NE area, where Greedy+ hardly delivers any messages into the Destination Area, while S-FoT+ --- using ten times as many messages --- is seventeen times more reliable. Detailed numbers are shown in Table~\ref{tbl:UrbanResults}.

\begin{figure}[tbh!]
    \centering
    \includegraphics[width=0.8\textwidth]{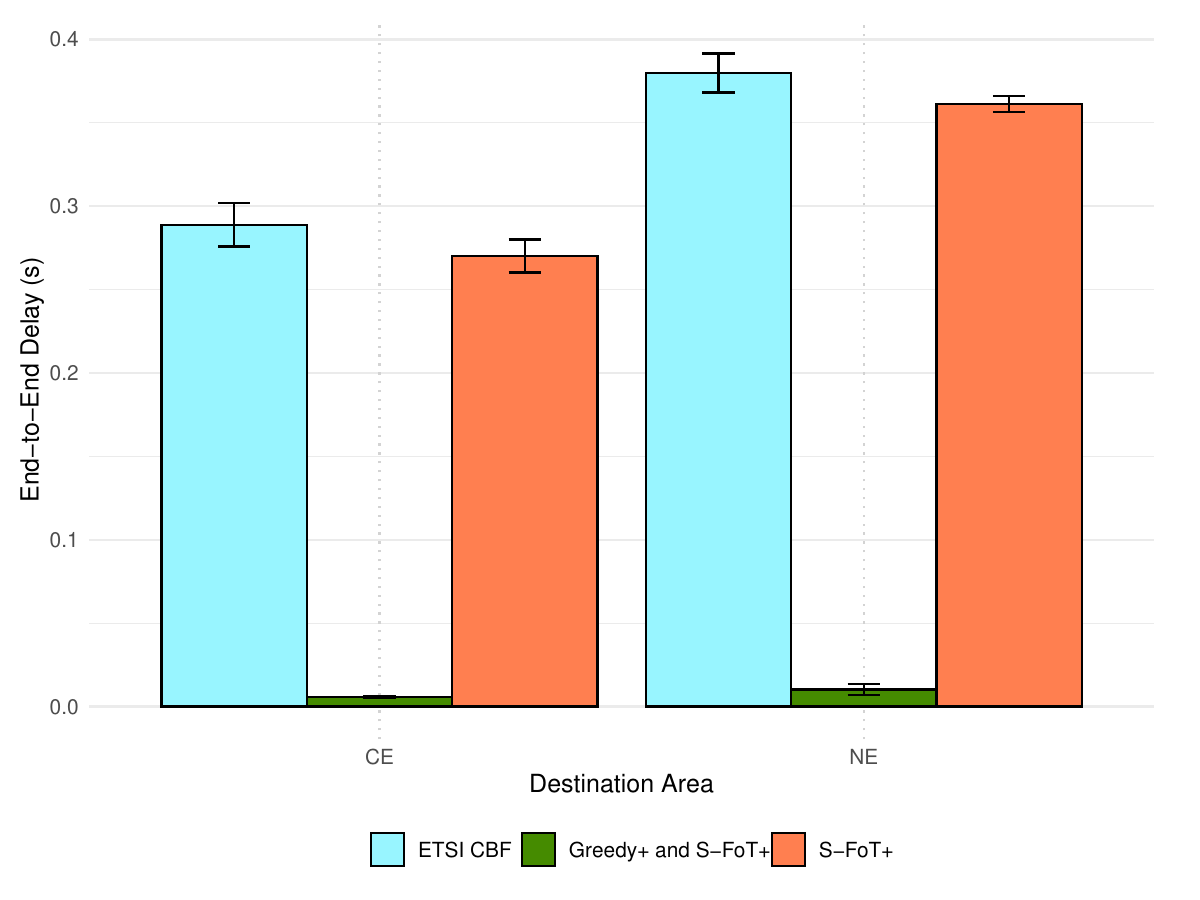}
    \caption{Average End-to-End delay for the first successful arrivals on each Destination Area}
    \label{fig:e2e_urban}
\end{figure}

Nevertheless, Greedy+ packets that do arrive in the area, do so in a significantly lower time, as shown on Fig.~\ref{fig:e2e_urban}, which records latencies for the first packet that arrives successfully in the Destination Area (again, \ac{PDR} has to be factored in when interpreting this figure). Values for end-to-end delay follow an interesting trend: even if the NE Destination Area is practically twice as far as the CE area, latency is not twice as large, which can be attributed to the path a message follows through the streets toward its destination.

\begin{figure}[p!]
    \centering
    \includegraphics[width=1.0\textwidth]{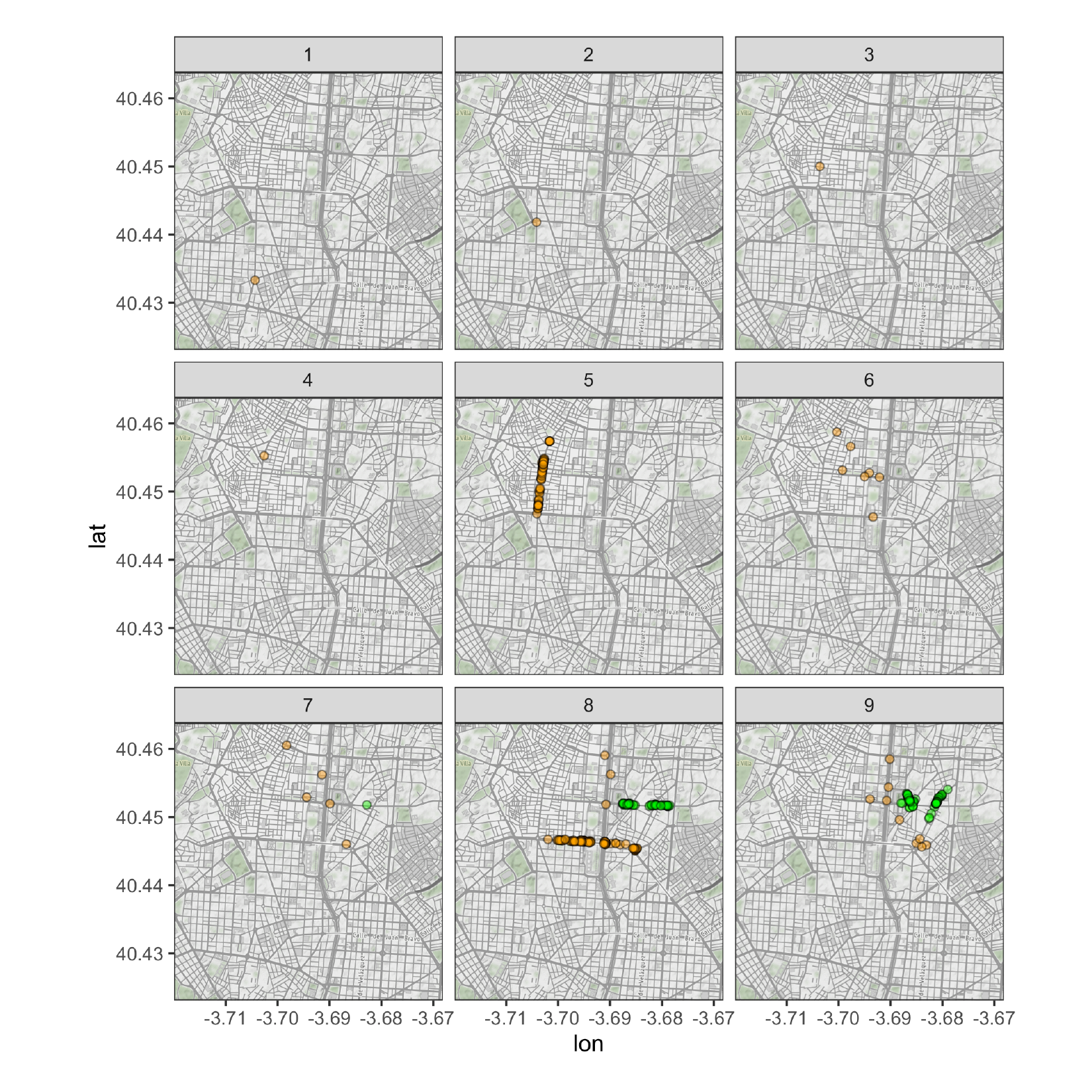}
    \caption{Greedy+ dissemination toward the northeast Destination Area}
    \label{fig:greedy_hoptrack_10}
\end{figure}

Fig.~\ref{fig:greedy_hoptrack_10} shows the path a message follows toward the NE area using Greedy+. Every subfigure is labeled with the hop number at which the packet was buffered ("hop 0" is the original message from the source). It can be observed that, even if a straight, diagonal path from the southwest to the northeast is the shortest way, the message favors going along streets. Until the fourth hop, the packet travels north until the forwarder determines it is a local maximum and decides to broadcast the packet to scout for a different path (hop 5). From there, multiple forwarders select next hops until the first message arrives at the Destination Area on hop 7, with only two remaining hops to live (the maximum number of hops is set to 10 in this simulation), and not managing to be disseminated to every node in the area. This prompts the question of whether an extended \ac{TTL} may benefit forwarding.

\begin{figure}[tbh!]
    \centering
    \includegraphics[width=0.7\textwidth]{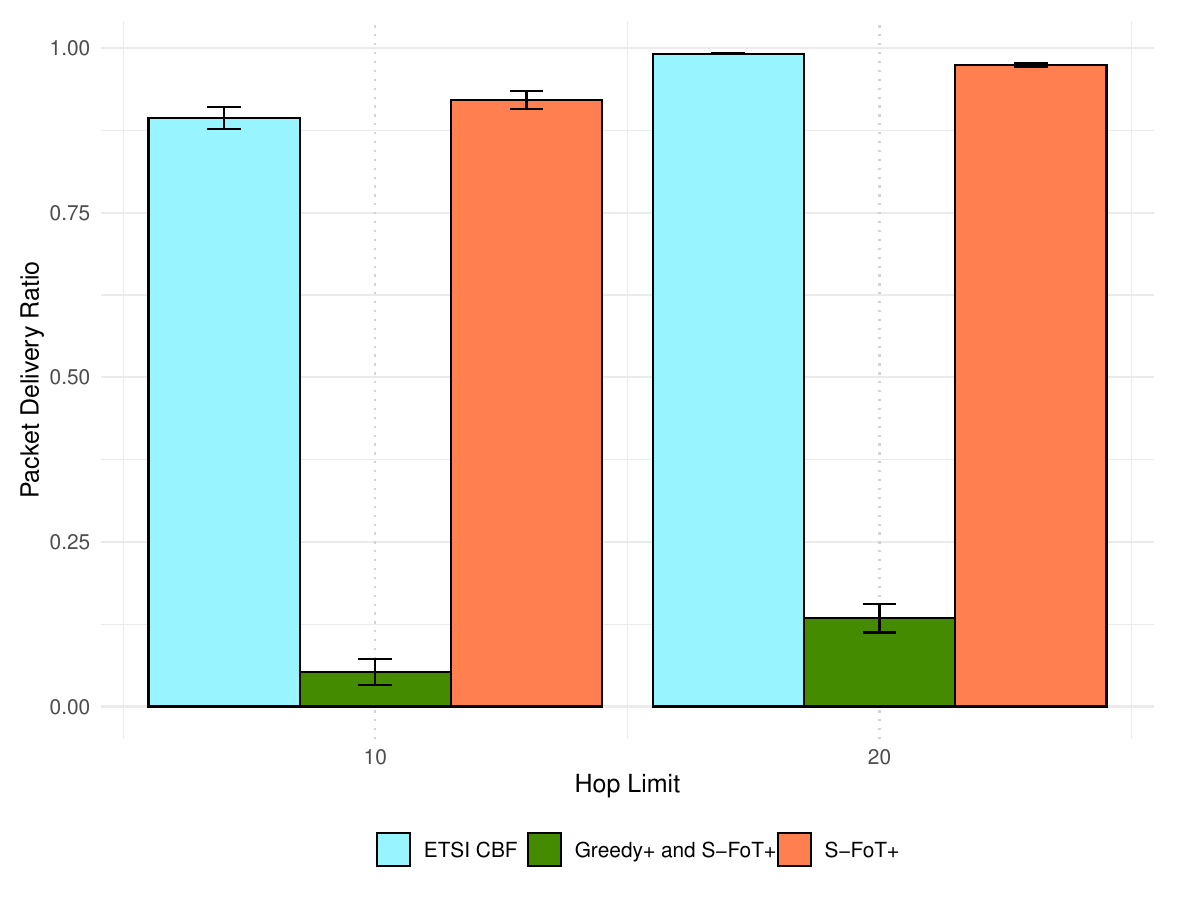}
    \caption{Packet-delivery Ratio comparison with two hop limits for the farthest Destination Area (NE)}
    \label{fig:pdr_urban_hc20}
\end{figure}

\begin{figure}[tbh!]
    \centering
    \includegraphics[width=0.7\textwidth]{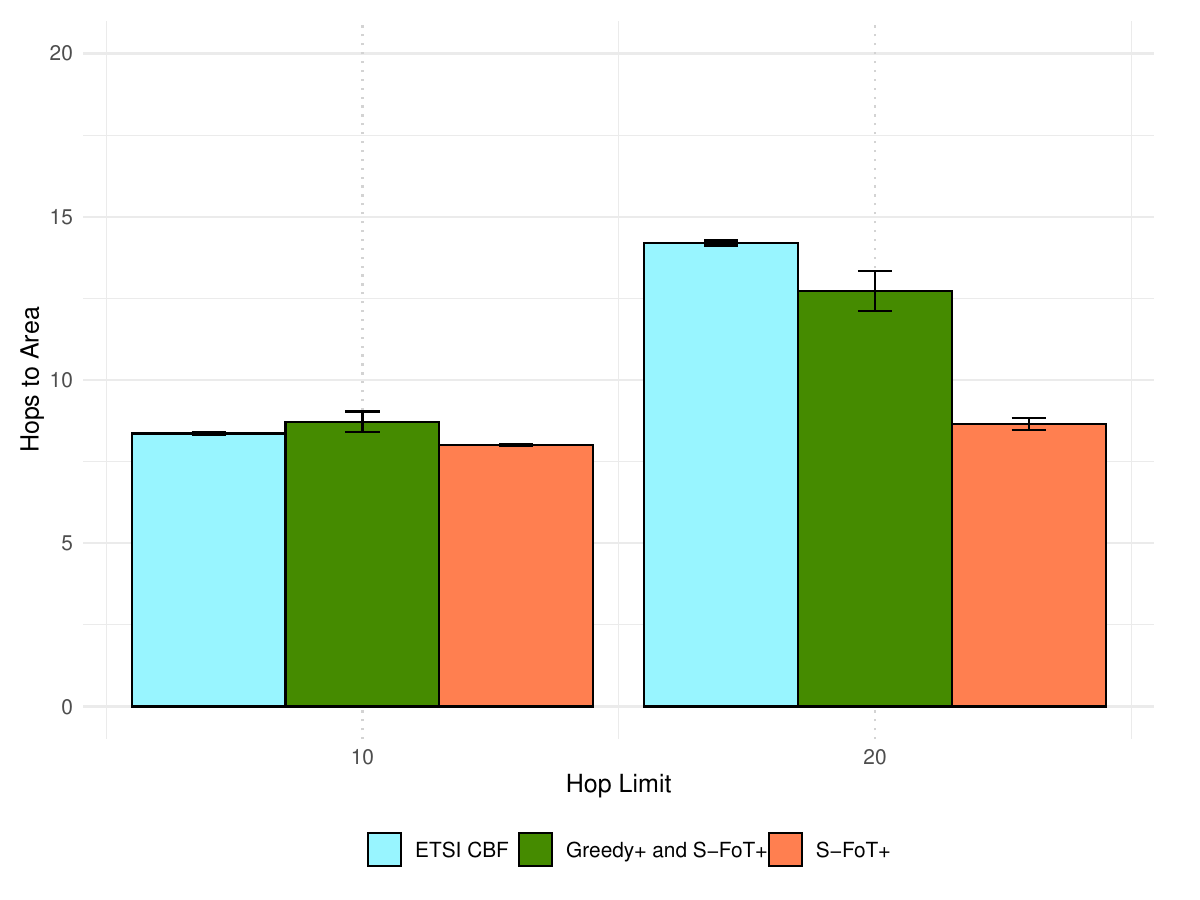}
    \caption{Average number of hops to reach vehicles in the Destination Area}
    \label{fig:hops_urban}
\end{figure}

In order to address this question, we have evaluated the results for the northeast (NE) area with an increased hop limit (\ac{TTL}). We have increased the \ac{TTL} value from 10 to 20 and repeated the experiment. Fig.~\ref{fig:pdr_urban_hc20} shows the success rate for the two values. Even though there is an improvement for Greedy+, it is still clearly outperformed by purely receiver-based approaches. ETSI CBF using the extended \ac{TTL} even achieves slightly better \ac{PDR} results than S-FoT+. Fig.~\ref{fig:hops_urban} shows the average hop at which a packet is buffered by nodes inside the area. Both ETSI CBF and Greedy+ are prone to "hop exhaustion", and use more hops when offered to. S-FoT+, however, uses, in average, less than an extra hop when the hop limit is increased in 10 hops.

\begin{figure}[p!]
    \centering
    \includegraphics[width=1.0\textwidth]{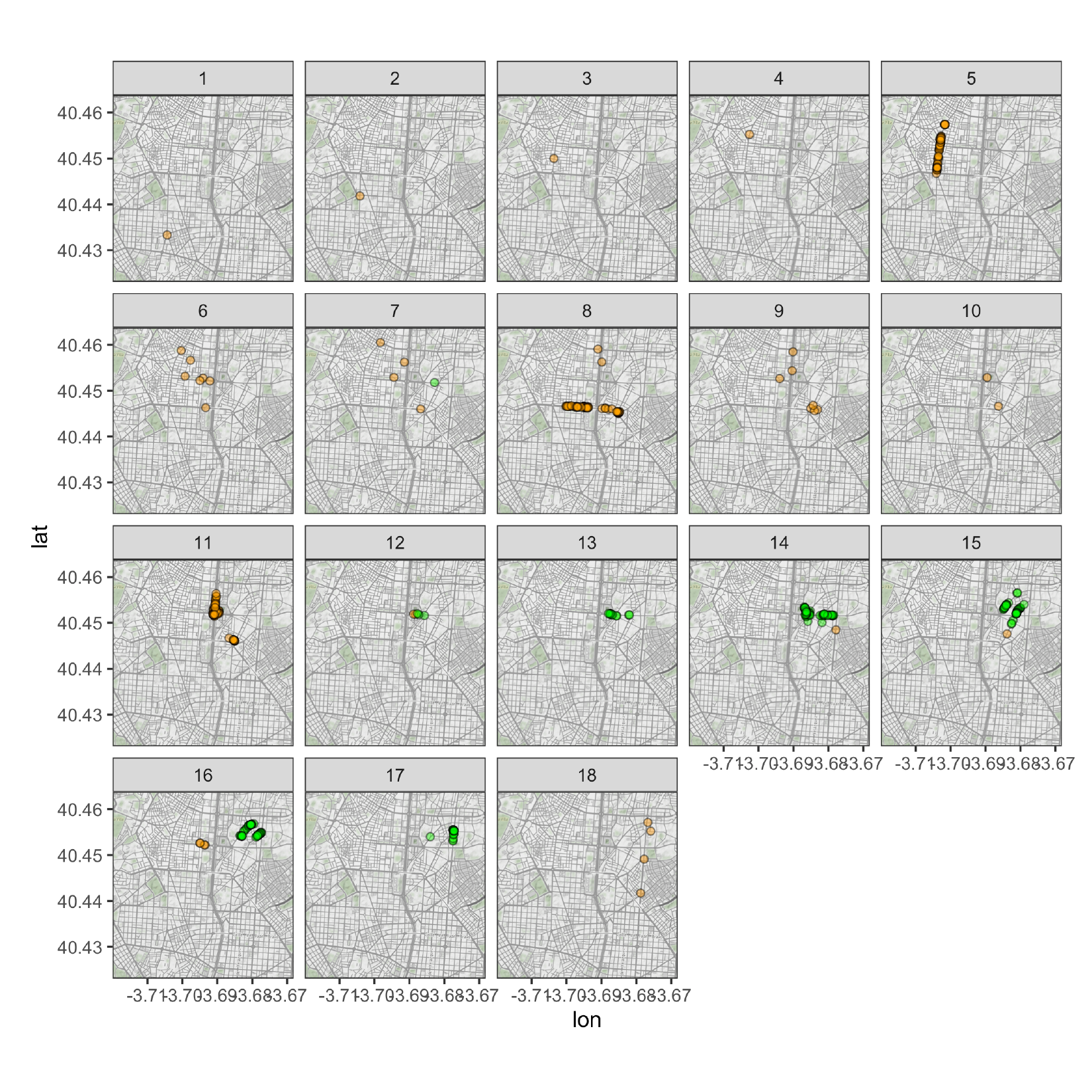}
    \caption{Greedy+ dissemination with an increased hop limit (20)}
    \label{fig:greedy_hc20}
\end{figure}

\begin{figure}[p!]
    \centering
    \includegraphics[width=1.0\textwidth]{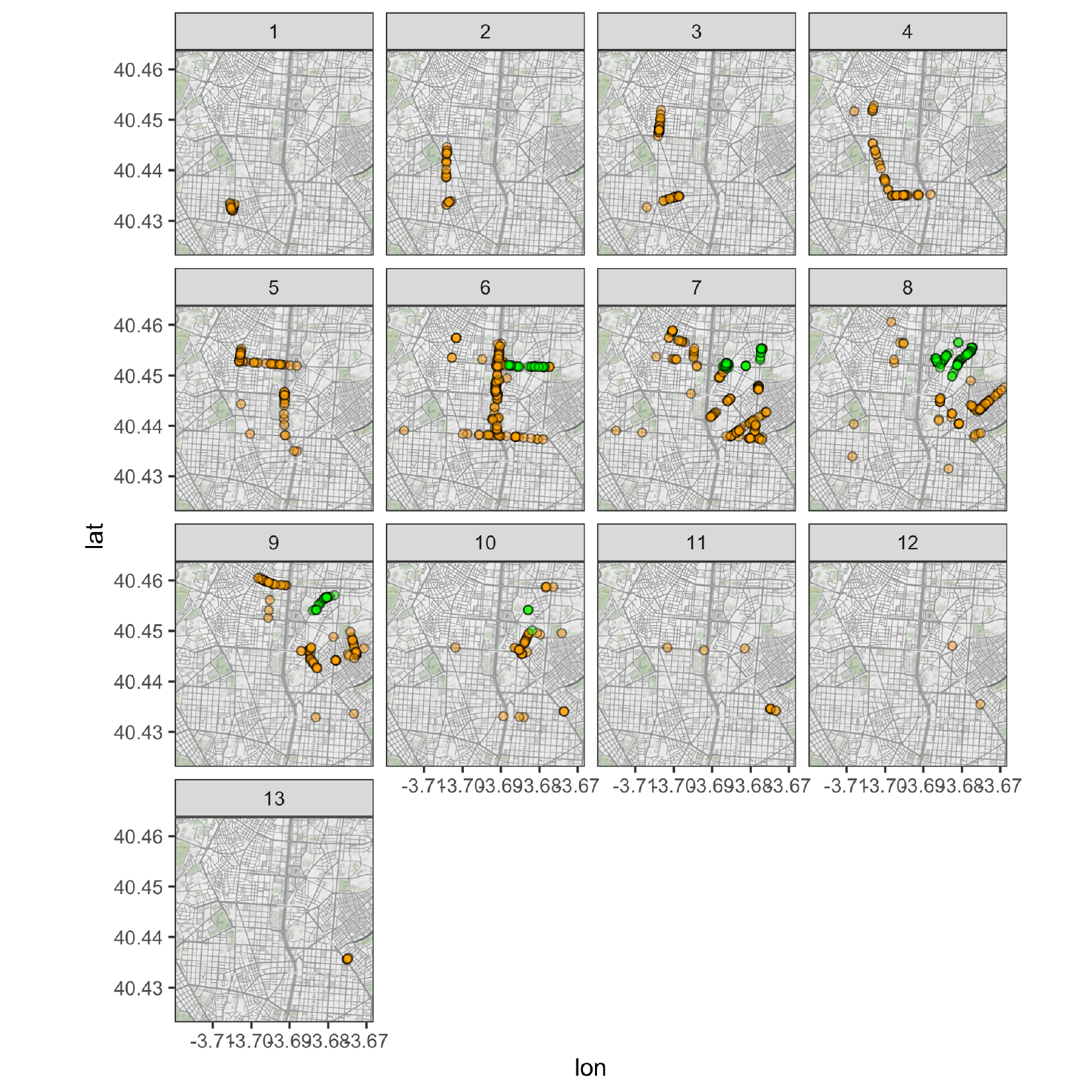}
    \caption{S-FoT+  dissemination with an increased hop limit (20)}
    \label{fig:sfot_hoptrack}
\end{figure}

\begin{figure}[p!]
    \centering
    \includegraphics[width=1.0\textwidth]{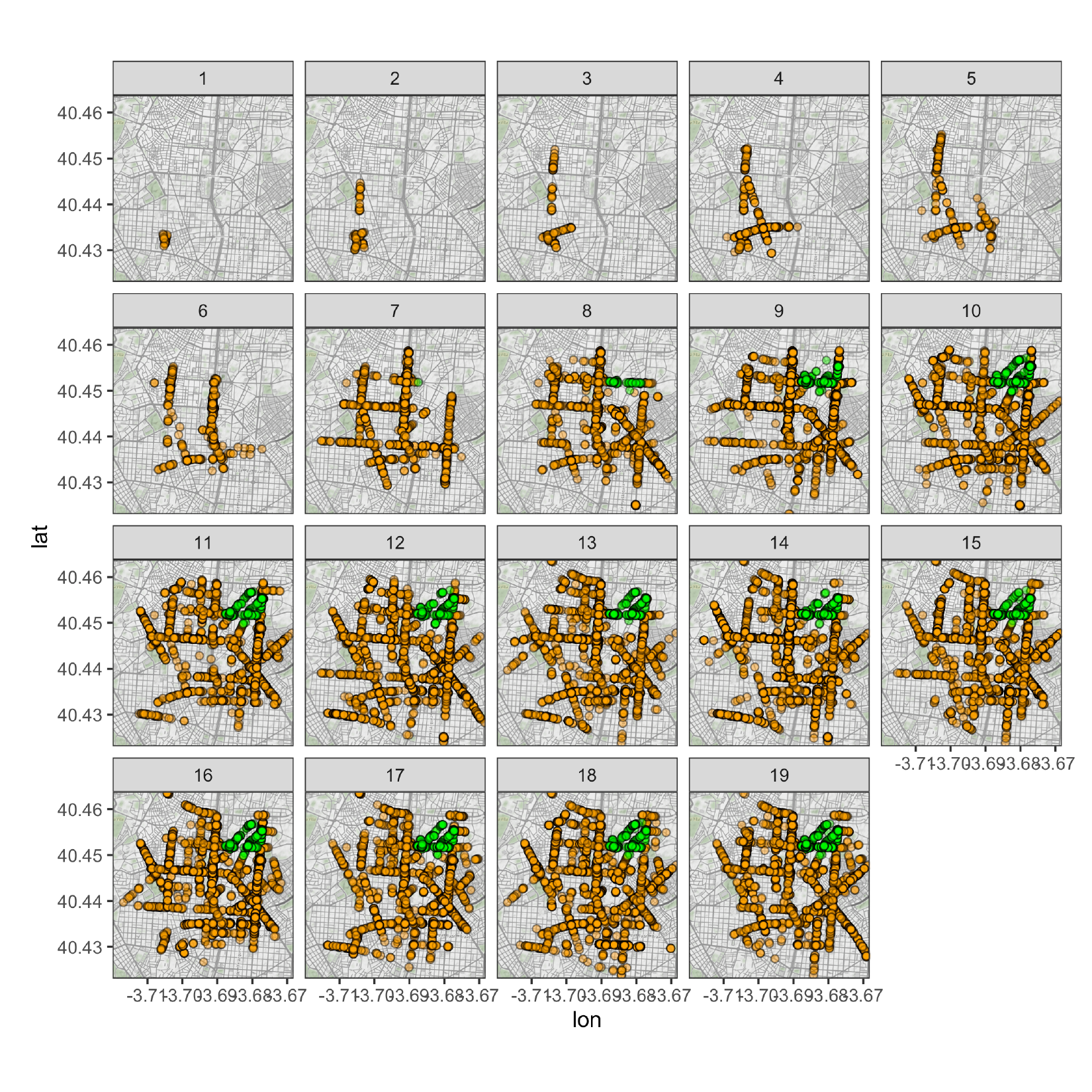}
    \caption{ETSI CBF dissemination with an increased hop limit (20)}
    \label{fig:cbf_hoptrack}
\end{figure}

\begin{figure}[tbh!]
    \centering
    \includegraphics[width=0.8\textwidth]{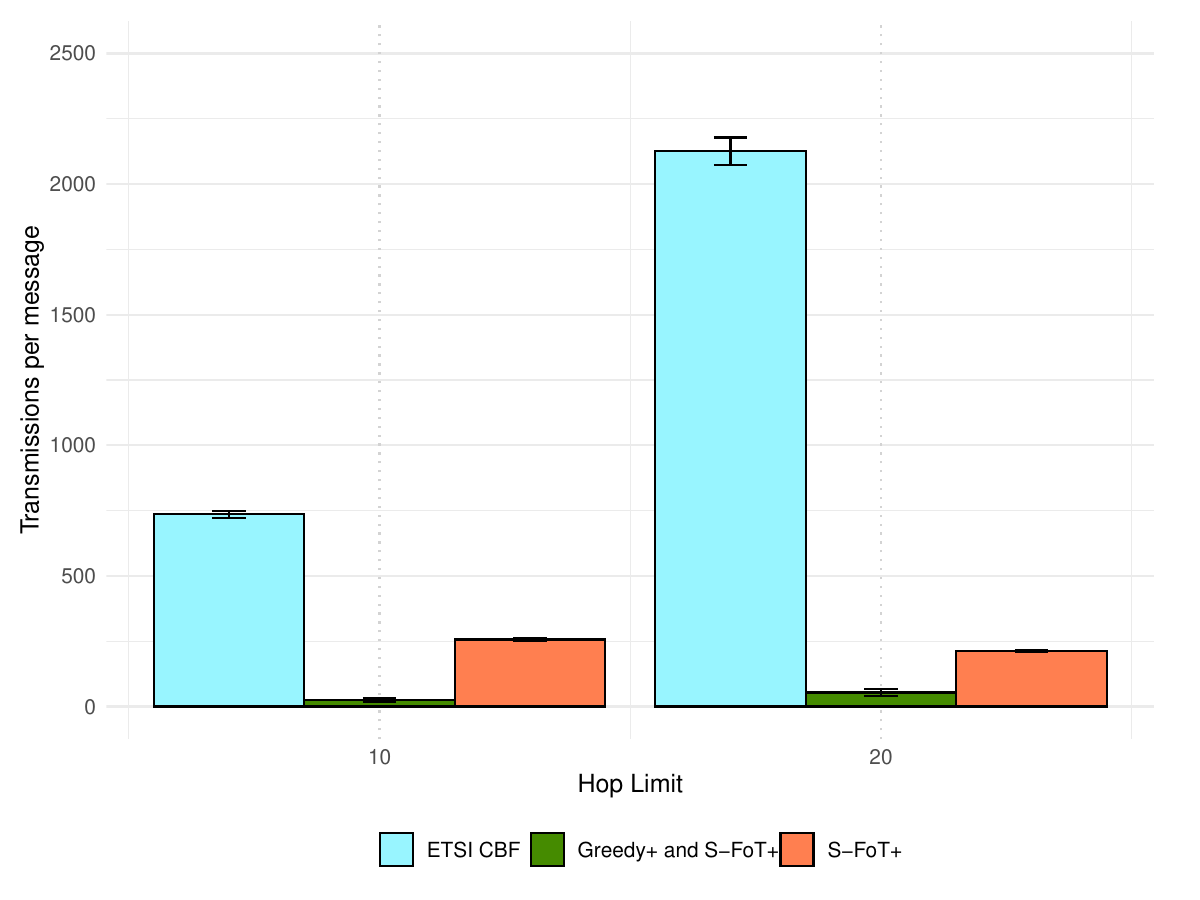}
    \caption{Transmission comparison for two hop limits}
    \label{fig:txd_hc}
\end{figure}

Fig.~\ref{fig:greedy_hc20} shows the dissemination pattern for one message using Greedy+ combined with S-FoT+ with an increased \ac{TTL}. As Greedy+ does, next-hop determination favors neighbors along streets, but the broadcast nature of S-FoT+ allows for the scouting of parallel paths. On the one hand, Fig.~\ref{fig:greedy_hc20} shows that Greedy+ needs more hops to even reach the Destination Area. On the other hand, Fig.~\ref{fig:sfot_hoptrack} shows that S-FoT+ uses its \ac{TTL} efficiently and only requires one extra hop to reach some additional stations in the Destination Area (note that the original hop limit of 10 includes hops 1 to 9 in Fig.~\ref{fig:sfot_hoptrack}). However, these extra hops also give ETSI CBF more room for inefficiency. Fig.~\ref{fig:cbf_hoptrack} shows the dissemination pattern for ETSI CBF with an extended \ac{TTL}, In this case, even when the Destination Area is effectively covered by hop 10, the message is still around and being re-transmitted due to the lack of \ac{DPD} in ETSI CBF. It is only due to the message exhausting its \ac{TTL} that the broadcast storm stops. The consequence of this phenomenon is reflected on Fig.~\ref{fig:txd_hc}, which shows that transmissions per message skyrocket for ETSI CBF while they are even reduced for S-FoT+, since the extra hops allow for unnecessary forwarding of packets to be canceled.

The results from the evaluation in urban settings concur with the results from Section~\ref{sec:highway} in that Greedy Forwarding is significantly less reliable than receiver-based mechanisms, and that the trade-off in latency cannot override the need to disseminate information effectively. The additional challenges that come with urban settings affect Greedy Forwarding more than they do receiver-based algorithms, which are more effective (e.g., in terms of \ac{PDR}). Among these two, S-FoT+ reaches these efficacy goals in a more efficient way (e.g., with only a fraction of the transmissions used by ETSI CBF).

\section{Related Work} 
\label{sec:relatedWork}
In this section we discuss other works that have focused on ETSI non-area forwarding algorithms, both Greedy Forwarding and Non-Area \ac{CBF}.

ETSI GeoUnicast service based on Greedy Forwarding (i.e., very similar to non-area forwarding, but with messages addressed to a single destination vehicle) is analyzed in \cite{Tao2017}. The authors propose to improve the \ac{PDR} by incorporating a multipath greedy scheme, where each message is forwarded to the destination through a number of paths (i.e., each sender transmits the message to the two or three best next hops depending the relevance of the message). Although this work considers the ETSI architecture, neither \ac{DCC} nor high vehicle density scenarios are taken into account. On the other hand, it should be noted that the problem of Greedy operation in congested channels (e.g., delay or starvation in DCC queues) that leads to poor \ac{PDR} is the same for each single path of the multipath solution.

In \cite{xu:2017}, it is proposed to enhance the ETSI Greedy Forwarding by taking into account the Received Signal strength Indicator (RSSI) observed between neighbors when transmitting packets toward the Destination Area. Additionally, the contention timer of the Area CBF algorithm is modified to consider the heading of the vehicles. Both modifications of the ETSI algorithms are evaluated in urban real field tests with a small Destination Area of $80\,m^2$, and a distance from the source vehicle to the centre of the area of 276\,m. The conclusions do not report any particular findings on Greedy Forwarding. Moreover, the scope of the results is rather limited considering the low density of vehicles (around 16) and the low traffic on the network.

On the other hand, the use of Non-Area CBF is evaluated in \cite{Paulin2015} where two main inefficiencies are identified, namely, the need for persistent \acl{DPD}, and the case where two forwarders in close proximity transmit at the same time (without canceling each other out) causing all candidate forwarders to be canceled, which stops the forwarding of the packet. For the second issue, a randomized cancellation of packets in the CBF buffer is proposed, so that the reception of a packet that is already buffered does not always cause its discarding (which also leads to more transmissions). However, neither the delivery to the application layer nor the complete ETSI stack architecture is taken into account (the congestion control mechanism is not considered), and the performance evaluation is carried out by simulation using only static vehicles on highways. It should be noted that S-FoT+ also includes a mechanism to restrict packet cancellation in the CBF buffer. The S-FoT+ mechanism instead of being aleatory takes into account the progress toward the destination, which is more accurate.

\section{Conclusions and Future Work}
\label{sec:conclusions-fw}

We presented an evaluation of ETSI GeoNetworking for scenarios where a message has to be received in a Destination Area when the source node is a distant node, located outside of this area. The ETSI-defined mechanisms to route a message toward a Destination Area are sender-based Greedy Forwarding and receiver-based Non-Area \ac{CBF}. We also compared ETSI-defined mechanisms to several optimizations found in the literature. For Greedy Forwarding, we have chosen optimizations that allow the selection of a better next hop by limiting the sense of neighborhood in time and space. For Non-Area \ac{CBF}, we have chosen optimizations that improved efficiency in Area \ac{CBF} and adapted them to Non-Area \ac{CBF} scenarios.

The first takeaway from this work is that the GeoNetworking combination suggested by ETSI in~\cite{etsiNewGeoNetworking} (i.e., Greedy Forwarding outside of the Destination Area and CBF in the inside) is ineffective. This is due to the inability of Greedy Forwarding to reach a remote Destination Area. While the text in Annex F of~\cite{etsiNewGeoNetworking} acknowledges Greedy has reliability issues, we show that unless optimizations are used, its reliability with \ac{DCC} is null. Furthermore, even when these optimizations improve reliability by enabling Greedy to select a better candidate for the next hop, we have also found that efficacy is very poor when channel congestion is present (i.e., in high vehicle density scenarios) because the interaction with DCC. This is due to the nature of sender-based mechanisms, where a candidate is selected as the next hop regardless of its actual ability to forward a message.

Secondly, \ac{CBF} is much more reliable by offering more paths for a message to follow toward the Destination Area. Receiver-based mechanisms provide a fallback for cases where the optimal candidate is unable to forward a message. ETSI Non-Area CBF outperforms Greedy Forwarding both in highway and urban scenarios. However, due to its proven inefficiency, it does so at the cost of using a significant number of resources (i.e., transmissions).

The last takeaway from this work refers to the efficiency problem. While previous works had shown that S-FoT+ outperforms ETSI CBF in area forwarding, we have now assessed its performance in non-area scenarios. Results show that S-FoT+ maintains its advantages over ETSI CBF also in non-area forwarding, by keeping \ac{PDR} as high, while drastically reducing the number of transmissions.

Furthermore, even if Greedy Forwarding has the ability to reach the Destination Area with a very low latency, its shortcomings in PDR render this advantage useless. The effect of different phenomena, e.g., congestion and the sense of neighborhood, makes \ac{CBF} more effective, especially S-FoT+ which is aware of channel congestion. In summary, using S-FoT+ both within and outside the Destination Area exhibits the best performance both in efficacy and efficiency.

Future work will be performed in assessing if and how the identified phenomena affects other medium access technologies, like 5G~NR. In this paper, we have identified problems stemming from the effect of the Network \& Transport layer sending a packet to the Access layer assuming it will be transmitted almost immediately only to be stopped by DCC in ITS-G5. A similar effect can occur in 5G~NR if the Network \& Transport layer is not aware of the status of the Access layer, e.g., if a packet has to wait for the radio resources assigned by the scheduler.

\bibliography{mybibfile}

\end{document}